\documentclass[11pt,letterpaper]{article}

\usepackage{def}
\usepackage[retainorgcmds]{IEEEtrantools}
\usepackage{amssymb}
\usepackage{amsmath}
\usepackage{multirow}

\font\ro=cmsy10                          % font with rope
\def\kcr{{\hbox{\ro \char'170}}}                % right-handed rope
\def\ktl{{\hbox{\ro \char'170}}}        % top end for left-handed rope
\def\ktr{{\hbox{\ro \char'170}}}        % " right
\def\kbl{{\hbox{\ro \char'170}}}        % " bottom left
\def\kbr{{\hbox{\ro \char'170}}}        % " right
\parskip=\medskipamount                 % space between paragraphs (LaTeX)
\thicklines                         % thick straight lines for pictures (LaTeX)

\def\border{                                            % border
        \setlength{\unitlength}{1mm}
        \newcount\xco
        \newcount\yco
        \xco=-21
        \yco=12
        \begin{picture}(140,0)
        \put(\xco,\yco){$\ktl$}
        \advance\yco by-1
        {\loop
        \put(\xco,\yco){$\kcr$}
        \advance\yco by-2
        \ifnum\yco>-240
        \repeat
        \put(\xco,\yco){$\kbl$}}
        \xco=158
        \yco=12
        \put(\xco,\yco){$\ktr$}
        \advance\yco by-1
        {\loop
        \put(\xco,\yco){$\kcr$}
        \advance\yco by-2
        \ifnum\yco>-240
        \repeat
        \put(\xco,\yco){$\kbr$}}
        \put(-20,13){\tiny **University of Maryland * Center for String and
         Particle  Theory* Physics Department***University of Maryland *Center
        for String and Particle  Theory** }
        \put(-20,-241.5){\tiny **University of Maryland * Center for String and
         Particle  Theory* Physics Department***University of Maryland *Center
        for String and Particle  Theory** }
        \end{picture}
        \par\vskip-8mm}

\def\headpic{                                           % UM heading
        \indent
        \setlength{\unitlength}{.4mm}
        \thinlines
        \par
        \begin{picture}(29,16)
        \put(165,16){\line(1,0){4}}
        \put(170,16){\line(1,0){4}}
        \put(180,16){\line(1,0){4}}
        \put(175,0){\line(1,0){4}}
        \put(180,0){\line(1,0){4}}
        \put(185,0){\line(1,0){4}}
        \put(169,0){\line(0,1){16}}
        \put(170,0){\line(0,1){16}}
        \put(179,0){\line(0,1){16}}
        \put(180,0){\line(0,1){16}}
        \put(184,0){\line(0,1){16}}
        \put(185,0){\line(0,1){16}}
        \put(169,16){\oval(8,32)[bl]}
        \put(170,16){\oval(8,32)[br]}
        \put(179,0){\oval(8,32)[tl]}
        \put(185,0){\oval(8,32)[tr]}
        \end{picture}
        \par\vskip-6.5mm
        \thicklines}
\def\endtitle{\end{quotation}\newpage}                  % end title page

\begin{document}

\border\headpic {\hbox to\hsize{August 2013 \hfill
{UMDEPP-013-016}}}
\par
{$~$ \hfill
{hep-th/1310.7386}
}
\par

\setlength{\oddsidemargin}{0.3in}
\setlength{\evensidemargin}{-0.3in}
\begin{center}
\vglue .10in
{\large\bf On 4D, ${\cal N} = 1$ Massless Gauge Superfields of\\
Higher Superspin: Half-Odd-Integer Case
 \footnote
{Supported in part  by National Science Foundation Grant
PHY-09-68854.}\  }
\\[.5in]

S.\, James Gates, Jr.\footnote{gatess@wam.umd.edu}
and
Konstantinos Koutrolikos\footnote{koutrol@umd.edu}
\\[0.2in]

{\it Center for String and Particle Theory\\
Department of Physics, University of Maryland\\
College Park, MD 20742-4111 USA}\\[1.8in]

{\bf ABSTRACT}\\[.01in]
\end{center}
\begin{quotation}
{We present an alternative method to explore the off-shell 
component structure of theories that describe half-integer 
super-helicities $Y=s+1/2$ (where $s$ is any positive integer). 
We use it to derive the component action, component SUSY 
transformation laws, and count the component-level degrees 
of freedom involved. This counting will give us clues about 
$\mathcal{N}=2$ representations. The foundation of the 
process relies on the superfield equations of motion, generated 
by variation of a superspace action expressed in terms of 
prepotentials.  Using this approach we reproduce the half-integer 
super-helicity superspace action using unconstrained superfields.
}

%${~~~}$ \newline
%PACS: 04.65.+e
\endtitle
 
\section{Introduction}
~~ In the preceding paper \cite{IntSpin}
we discussed the case of integer super-helicity theories. In this 
complementary paper, the corresponding program is carried out 
for the case of half-integer super-helicities. Following the same 
strategy as in \cite{IntSpin}, we use representation theory as a 
guideline to dictate the proper type of superfields we should 
consider for the construction of the theory. 

Under our restriction\footnote{The formalism in \cite{KSP} does 
permit violations of this condition.} using auxiliary superfields 
with lower spins than the main gauge superfield, unlike the 
integer case, we receover two different formulations for the 
description of the highest possible super-helicity, in agreement 
with the results in \cite{KSP}.  We will verify that although they 
describe the same physical system on-shell, they don't have 
the same off-shell structure and they involve different numbers 
of degrees of freedom. From that point of view, we can say that 
these two theories are not equivalent off-shell, meaning that 
there is no 1-1 mapping between the two.

After the construction of the superspace action in terms of unconstrained 
superfields, we use the equations of motion and their properties, 
such as the Bianchi identities, to define the various components, 
derive the component action and their SUSY-transformation laws. 
We also do a counting of the off-shell degrees of freedom. A simple 
counting argument provides a supporting case for pairs of $\mathcal{N}
=1$ theories noted before  \cite{GKS} to create $\mathcal{N}=2$ 
irreducible higher spin representations. 

This paper is organized as follows: In section \ref{Irreps} we 
quickly review the representation theory of the little group of 
the $4D,~\mathcal{N}=1$ Super-Poincar\'{e} group for a 
half-integer superspin/helicity system. In section \ref{Redndcy} 
we focus on the massless case and illustrate how the gauge 
transformation of the superfield emerges. In section \ref{Action}, 
using the invariance of the physical degrees of freedom as a 
guideline, we build the superspace action of the theory and 
prove that it describes the desired super-helicity. As mentioned 
before there are two ways to do this and we will present both. The 
next section \ref{Projection} we discuss the off-shell components 
for both of these theories. Using the equations of motion of the 
superspace action we define the off-shell components, obtain 
the component action in a diagonal form and explicit expressions 
for their SUSY-transformation laws.

\section{Irreducible Representations}
\label{Irreps}
~~ Following the review of the representation theory presented 
in \cite{IntSpin},\cite{BK} we conclude that the two cases of massive 
and massless have the following properties discussed below.

\subsection{Massive Case}
~~ For the massive case, the superfield which describes a real 
irreducible representation of half-integer superspin $Y=s+1/2$ 
(and it is the highest superspin that it can describe) is a real 
bosonic superfield $H_{\a(s)\ad(s)}$ with $s$ undotted 
symmetrized indices and $s$ dotted symmmetrized indices 
and must satisfy the constraints
\bea{l}
\D^2 H_{\a(s)\ad(s)}=0\\
\Dd^2 H_{\a(s)\ad(s)}=0\\
\D^{\g}H_{\g\a(s-1)\ad(s)}=0\IEEEyesnumber\\
\pa^{\g\gd}\Phi_{\g\a(n-1)\gd\ad(m-1)}=0\\
\Box H_{\a(s)\ad(s)}=m^2 H_{\a(s)\ad(s)}
\eea
Equivalently there is a chiral superfield $W_{\a(s+1)\ad(s)}$ defined 
as
\be
W_{\a(s+1)\ad(s)}=\frac{1}{(s+1)!}\Dd^2\D_{(\a_{s+1}}H_{\a(s))\ad(s)}\\
\ee
with
\bea{l}
\Dd_{\bd}W_{\g\a(s)\ad(s)}=0,\ \text{chiral}\\
\pa^{\b\bd}W_{\b\a(s)\bd\ad(s-1)}=0\IEEEyesnumber\\
\Box W_{\a(s+1)\ad(s)}=m^2W_{\a(s+1)\ad(s)}
\eea
The spin content of this supermultiplet is $j=s+1,\ s+1/2,\ s+1/2,\ s$.

\subsection{Massless Case}
~~ For the massless case, the half-integer super-helicity representation 
is described by a chiral superfield $F_{\a(2s+1)}$ with $2s+1$ 
symmetrized undotted indices and no dotted indices. It must satisfy 
the constraints
\bea{l}
\Dd_{\gd}F_{\a(2s+1)}=0,\ \text{chiral}\\
\D^{\b}F_{\b\a(2s)}=0\IEEEyesnumber\\
\eea
and the helicity content is $h=s+1,\ s+1/2$

\section{Massless limit and Redundancy}
\label{Redndcy}
~~ Now that we know the proper building blocks for the two 
irreducible representations we impose the convenient feature 
that the massless limit of the massive representation gives the 
massless representation plus other sectors that decouple.

In order for something like this to occur, we should be able to 
construct $F_{\a(2s+1)}$ out of the remaining objects after the 
limit of the massive theory has been taken. Given the chirality 
properties of $F$ and $W$ and their index structure\footnote{
The sum of the indices of $W$ is the number of the undotted 
indices of $F$} we can guess a mapping that could do the 
trick.
\be
F_{\a(2s+1)}\sim\pa_{(\a_{2s+1}}{}^{\ad_s}\ldots\pa_{\a_{
s+2}}{}^{\ad_1}\Dd^2\D_{\a_{s+1}}H_{\a(s))\ad(s)}
\ee

As it is explained in \cite{IntSpin}
\footnote{There the argument was for integer super-helicities, 
but it can be repeated for the half-integer case}that identification 
is problematic because $F$ is the object that carries the physical 
gauge-invariant degrees of freedom and not $H$ and also the 
degrees of freedom of $F$ and $H$ don't match. The way out 
of that is to introduce a redundancy and identify $H_{\a(s)\ad(s)}$ 
with $H_{\a(s)\ad(s+)}+R_{\a(s)\ad(s)}$.

The redundancy has to respect the physical (propagating) degrees 
of freedom of $F$ and leave them unchanged. Hence
\bea{l}
\pa_{(\a_{2s+1}}{}^{\ad_s}\ldots\pa_{\a_{s+2}}{}^{\ad_1}\Dd^2
\D_{\a_{s+1}}R_{\a(s))\ad(s)}=0\IEEEyesnumber
\eea
The most general solution\footnote{$R$ must be real since $H$ 
is real} to this is
\bea{l}
R_{\a(s)\ad(s-1)}=\frac{1}{s!}\D_{(\a_{s}}\bar{L}_{\a(s-1))
\ad(s)}-\frac{1}{s!}\Dd_{(\ad_s}L_{\a(s)\ad(s-1))}\IEEEyesnumber
\eea
This redundancy will be the gauge transformation of the superfield $H$

\section{The Superspace Action}
\label{Action}
~~ Using the equivalency class characterized by $H$ and the 
redundancy $R$ we attempt to construct a superspace action that 
will describe the irreducible representation of half-integer super-helicity. 
For that $H$ must have mass dimension zero\footnote{Its highest spin 
component is a propagating boson} and the action must involve four 
covariant derivatives\footnote{The action must be quadratic to $H$ 
and dimensionless}.

The most general action is
\bea{ll}
S=\int d^8z&\ a_1 H^{\a(s)\ad(s)}\D^{\g}\Dd^2\D_{\g}H_{\a(s)\ad(s)}\\
&+a_2 H^{\a(s)\ad(s)}\left\{\D^2,\Dd^2\right\} H_{\a(s)\ad(s)}\\
&+a_3 H^{\a(s)\ad(s)}\D_{\a_s}\Dd^2\D^{\g}H_{\g\a(s-1)\ad(s)}+
c.c.\IEEEyesnumber\\
&+a_4 H^{\a(s)\ad(s)}\D_{\a_s}\Dd_{\ad_s}\D^{\g}\Dd^{\gd}
H_{\g\a(s-1)\gd\ad(s-1)}+c.c.
\eea

The goal is to have a gauge invariant action $\delta_{G}S=0$, meaning
 the action respects the equivalence between $H$ and and $H+R$ and 
 therefore the physical degrees of freedom described by that action are 
invariant (gauge invariance). The strategy to do this is to pick the free 
parameters in a special way. If this is not possible then we introduce 
auxiliary superfields, compensators and/or put constraints on the 
parameter $L$ of the redundancy (gauge parameter).  We demand 
the compensators introduced, if necessary, will not contain degrees 
of freedom with spin higher or equal to that  in the main gauge superfield, 
therefore they must have fewer indices than the main object $H$.

The deformation of the action is:
\bea{ll}
\delta_G S=\int & d^8z\left[(-2a_1+2\frac{s+1}{s}a_3+2a_4)\D^2\Dd_{
\ad_s}H^{\a(s)\ad(s)}\right.\\
&~~~~+\left.(-2a_3-\frac{s+1}{s}a_4)\D^{\a_s}\Dd_{\gd}\D_{\g}H^{\g\a(
s-1)\gd\ad(s-1)}\right]
\left(\Dd^2 L_{\a(s)\ad(s-1)}\right.\\
&~~~~~~~~~~~~~~~~~~~~~~~~~~~~~~~~~~~~~~~~~~~~~~~~~~~
~~~~~+\left.\D^{\a_{s+1}}\Lambda_{\a(s+1)\ad(s-1)}\right)\\
&+2a_2 H^{\a(s)\ad(s)}\D^2\Dd^2\D_{\a_s}\bar{L}_{\a(s-1)\ad(s)}
\IEEEyesnumber\\
&-2a_4\Dd_{\bd}\D_{\g}\Dd_{\gd}H^{\g\a(s-1)\bd\gd\ad(s-2)}\left[
\Dd^{\ad_{s-1}}\D^{\a_s}L_{\a(s)\ad(s-1)}\right.\\
&~~~~~~~~~~~~~~~~~~~~~~~~~~~~~~~~~~~~~~~~~+\frac{s-1}{s}
\D^{\a_s}\Dd^{\ad_{s-1}}L_{a(s)\ad(s-1)}\\
&~~~~~~~~~~~~~~~~~~~~~~~~~~~~~~~~~~~~~~~~~\left.+\Dd_{
\ad_{s-2}}J_{\a(s-1)\ad(s-3)}\right]\\
&+c.c.
\eea

Notice that because of the $\D$-algebra we have the freedom to add 
terms like\\ $\D^{\a_{s+1}}\Lambda_{\a(s+1)\ad(s-1)}$ and $\Dd_{\ad_{
s-2}}J_{\a(s-1)\ad(s-3)}$ which identically vanish and they don't effect 
the result.

Obviously we can not set the variation of the action to zero just by 
picking values for the $a$'s without setting them all to zero, but we 
can introduce compensators with proper mass dimensionality and 
index structure. There are two different ways to do that
\begin{itemize}
\item (I) Choose coefficients to kill the last two terms ($a_2=a_4
=0$) and introduce a compensator that cancels the first term
\item (II) Choose coefficients to kill the first two terms\\
($-2a_1+2\frac{s+1}{s}a_3+2a_4=0$, $-2a_3-\frac{s+1}{s}a_4$, 
$a_2=0$) and introduce a compensator to cancel the last term
\end{itemize}

These two different approaches will lead to the two different formulations 
of half-integer super-helicity, mentioned above.

\subsection{Case (I) - Transverse theory}
~~For case (I) we find
\bea{l}
a_2=a_4=0\\
\eea
\bea{ll}
\delta_G S=\int & d^8z\left[(-2a_1+2\frac{s+1}{s}a_3)\D^2\Dd_{\ad_s
}H^{\a(s)\ad(s)}\right.\\
&~~~~+\left.-2a_3\D^{\a_s}\Dd_{\gd}\D_{\g}H^{\g\a(s-1)\gd\ad(s-1)}
\right]
\left(\Dd^2 L_{\a(s)\ad(s-1)}\right.\IEEEyesnumber\\
&~~~~~~~~~~~~~~~~~~~~~~~~~~~~~~~~~~~~~~~~~~~~~~~~~
~~~~~~~+\left.\D^{\a_{s+1}}\Lambda_{\a(s+1)\ad(s-1)}\right)\\
\eea

This suggests us to introduce a fermionic compensator $\chi_{\a(s
)\ad(s-1)}$ which transforms like $\d_G\chi_{\a(s)\ad(s-1)}=\Dd^2
L_{\a(s)\ad(s-1)}+\D^{\a_{s+1}}\Lambda_{\a(s+1)\ad(s-1)}$.
So in order to obtain invariance we add to the action two new pieces: 
The coupling term of $H$ with $\chi$ and the kinetic energy terms 
for $\chi$. The full action takes the form
\bea{ll}
S=\int d^8z&\ a_1 H^{\a(s)\ad(s)}\D^{\g}\Dd^2\D_{\g}H_{\a(s)\ad(s)}\\
&+a_3 H^{\a(s)\ad(s)}\D_{\a_s}\Dd^2\D^{\g}H_{\g\a(s-1)\ad(s)}+c.c.\\
&-(2a_1-2\frac{s+1}{s}a_3)H^{\a(s)\ad(s)}\Dd_{\ad_s}\D^2\chi_{\a(s)
\ad(s-1)}+c.c.\\
&+2a_3 H^{\a(s)\ad(s)}\D_{\a_s}\Dd_{\ad_s}\D^{\g}\chi_{\g\a(s-1)\ad(
s-1)}+c.c.\\
&+b_1 \chi^{\a(s)\ad(s-1)}\D^2\chi_{\a(s)\ad(s-1)}+c.c.\IEEEyesnumber\\
&+b_2 \chi^{\a(s)\ad(s-1)}\Dd^2\chi_{\a(s)\ad(s-1)}+c.c.\\
&+b_3 \chi^{\a(s)\ad(s-1)}\Dd^{\ad_s}\D_{\a_s}\bar{\chi}_{\a(s-1)\ad(s)}\\
&+b_4 \chi^{\a(s)\ad(s-1)}\D_{\a_s}\Dd^{\ad_s}\bar{\chi}_{\a(s-1)\ad(s)}\\
\eea
and it has to be invariant under
\bea{ll}
\delta_G H_{\a(s)\ad(s)}&=\frac{1}{s!}\D_{(\a_{s}}\bar{L}_{\a(s-1))\ad(
s)}-\frac{1}{s!}\Dd_{(\ad_s}L_{\a(s)\ad(s-1))}\IEEEyessubnumber\\
\delta_G \chi_{\a(s)\ad(s-1)}&=\Dd^2L_{\a(s)\ad(s-1)}+\D^{\a_{s+1}}
\Lambda_{\a(s+1)\ad(s-1)}\IEEEyessubnumber
\eea

The equations of motion of the superfields are the variation of the 
action with respect the superfield
\bea{l}
T_{\a(s)\ad(s)}=\frac{\delta S}{\delta H^{\a(s)\ad(s)}},~G_{\a(s)\ad(
s-1)}=\frac{\delta S}{\delta \chi^{\a(s)\ad(s-1)}}\IEEEyesnumber
\eea
and the invariance of the action gives the following Bianchi Identities
\bea{l}
\Dd^{\ad_s}T_{\a(s)\ad(s)}-\Dd^2 G_{\a(s))\ad(s-1)}=0
\IEEEyessubnumber\\
\frac{1}{(s+1)!}\D_{(\a_{s+1}}G_{a(s))\ad(s-1)}=0
\IEEEyessubnumber
\eea
The Bianchi identities fix all the coefficients
\bea{ll}
a_3=0, ~~ & b_3=0\\
b_1=-\frac{s+1}{s}a_1, ~~ & b_4=2a_1\\
b_2=0
\eea
and the final form of the action is:
\bea{ll}
S=\int d^8z&\left\{\vphantom{\frac12} c~H^{\a(s)\ad(s)}\D^{\g}\Dd^2
\D_{\g}H_{\a(s)\ad(s)}\right.\\
&-2c~H^{\a(s)\ad(s)}\Dd_{\ad_s}\D^2\chi_{\a(s)\ad(s-1)}+c.c.\\
&-\frac{s+1}{s}c~\chi^{\a(s)\ad(s-1)}\D^2\chi_{\a(s)\ad(s-1)}+c.c.
\IEEEyesnumber\\
&+\left.\vphantom{\frac12}2c~\chi^{\a(s)\ad(s-1)}\D_{\a_s}
\Dd^{\ad_s}\bar{\chi}_{\a(s-1)\ad(s)}\right\}\\
\eea

The expressions for the equations of motion are:
\bea{ll}
\label{E.Q.I}
T_{\a(s)\ad(s)}&=2c\D^\g\Dd^2\D_\g H_{\a(s)\ad(s)}\\
&~~+\frac{2c}{s!}\left(\D_{(\a_s}\Dd^2\bar{\chi}_{\a(s-1))\ad(s)}-
\Dd_{(\ad_s}\D^2\chi_{\a(s)\ad(s-1))}\right)\IEEEyessubnumber\\
G_{\a(s)\ad(s-1)}&=-2c\D^2\Dd^{\ad_s}H_{\a(s)\ad(s)}-2c\frac{
s+1}{s}\D^2\chi_{\a(s)\ad(s-1)}\\
&~~+\frac{2c}{s!}\D_{(\a_s}\Dd^{\ad_s}\bar{\chi}_{\a(s-1))\ad(s)}
\IEEEyessubnumber
\eea
where $c$ is a free overall parameter that can be absorbed in 
the definition of the superfields but for the moment we'll leave 
it as it is and fix it later when we define the components. 

The above action is the same as the transversely-linear theory 
presented in \cite{KSP} if we solve the constraints and express 
it in terms of the prepotential, but now we have an alternate 
understanding why we have to consider these types of superfields 
in order to construct the action and why they have these gauge 
transformation. 

Before we do anything else we must first prove that indeed this 
action describes the desired representation. Using the equations 
of motion we can now prove that a chiral superfield $F_{\a(2s+1)}
$ exists and satisfies the following Bianchi identity
\bea{ll}
\D^{\a_{2s+1}}&F_{\a(2s+1)}=\\
&=\frac{1}{2c}\frac{1}{(2s)!}\pa_{(\a_{2s}}{}^{\ad_{s}}\ldots
\pa_{\a_{s+1}}{}^{\ad_1}T_{\a(s))\ad(s)}\\
&+\frac{i}{2c}\frac{s}{2s+1}\frac{B}{B+\Delta}\frac{1}{(2s)!}\D_{
(\a_{2s}}\Dd^2\pa_{\a_{2s-1}}{}^{\ad_{s-1}}\ldots\pa_{\a_{s+1
}}{}^{\ad_1}G_{\a(s))\ad(s-1)}\\
&+\frac{1}{2c}\frac{s}{2s+1}\frac{1}{(2s)!}\D_{(\a_{2s}}\pa_{
\a_{2s-1}}{}^{\ad_{s}}\ldots\pa_{\a_{s}}{}^{\ad_1}\bar{G}_{
\a(s-1))\ad(s)}\IEEEyesnumber\\
&+\frac{i}{2c}\frac{s}{2s+1}\frac{\Delta}{B+\Delta}\frac{1}{
(2s)!}\D_{(\a_{2s}}\Dd^{\ad_s}\pa_{\a_{2s-1}}{}^{\ad_{s-1
}}\ldots\pa_{\a_{s+1}}{}^{\ad_{1}}T_{\a(s))\ad(s)}
\eea
where
\bea{l}
F_{\a(2s+1)}=\frac{1}{(2s+1)!}\Dd^2\D_{(\a_{2s+1}}\pa_{\a_{
2s}}{}^{\ad_{s}}\ldots\pa_{\a_{s+1}}{}^{\ad_{1}}H_{\a(s))\ad(s)}
\eea
and that proves that on-shell where $T_{\a(s)\ad(s)}$ =  $G_{\a(s)
\ad(s-1)}$ = 0, we find the desired constraints to describe a 
super-helicity $Y=s+1/2$ system. The constants $B$ and 
$\Delta$ are only constrained by $B+\Delta\neq 0$.

Like in the integer super-helicity case, this action and superfield 
configuration are not unique, but a simple representative of a 
two parameter family of equivalent theories.  To see that we perform
redefinitions of the superfields. Dimensionality and index structure 
allow us to do the following redefinition of $\chi$
\bea{l}
\chi_{\a(s)\ad(s-1)}\rightarrow\chi_{\a(s)\ad(s-1)}+z\Dd^{\ad_s}
H_{\a(s)\ad(s)}\IEEEyesnumber
\eea
where $z$ is a complex parameter. This operation will generate 
an entire class of actions and transformation laws which all are 
related by the above redefinition.

The generalized action is
\bea{ll}
S=\int d^8w&\ c~H^{\a(s)\ad(s)}\D^{\g}\Dd^2\D_{\g}H_{\a(s)\ad(
s)}\\
&-2c\left[1+\frac{s+1}{s}z\right]~H^{\a(s)\ad(s)}\Dd_{\ad_s}\D^2
\chi_{\a(s)\ad(s-1)}+c.c.\\
&-2c\bar{z}~H^{\a(s)\ad(s)}\D_{\a_s}\Dd_{\ad_s}\D^{\g}\chi_{\g
\a(s-1)\ad(s-1)}+c.c.\\
&-2c\bar{z}\left[1+\frac{s+1}{s}\bar{z}\right]~H^{\a(s)\ad(s)}\D_{
\a_s}\Dd^2\D^{\g}H_{\g\a(s-1)\ad(s)}+c.c.\IEEEyesnumber\\
&-c|z|^2~H^{\a(s)\ad(s)}\D_{\a_s}\Dd_{\ad_s}\D^{\g}\Dd^{\gd}
H_{\g\a(s-1)\gd\ad(s-1)}+c.c.\\
&-\frac{s+1}{s}c~\chi^{\a(s)\ad(s-1)}\D^2\chi_{\a(s)\ad(s-1)}+c.c.\\
&+2c~\chi^{\a(s)\ad(s-1)}\D_{\a_s}\Dd^{\ad_s}\bar{\chi}_{\a(s-1)
\ad(s)}\\
\eea
and the generalized transformation laws are
\bea{ll}
\delta_G H_{\a(s)\ad(s)}&=\frac{1}{s!}\D_{(\a_{s}}\bar{L}_{\a(s-1))
\ad(s)}-\frac{1}{s!}\Dd_{(\ad_s}L_{\a(s)\ad(s-1))}
\IEEEyessubnumber\\
\delta_G \chi_{\a(s)\ad(s-1)}&=\left[1+\frac{s+1}{s}z\right]\Dd^2
L_{\a(s)\ad(s-1)}-\frac{z}{s!}\Dd^{\ad_s}\D_{(\a_s}\bar{L}_{\a(s-1
))\ad(s)}\IEEEyessubnumber\\
&~~+\D^{\a_{s+1}}\Lambda_{\a(s+1)\ad(s-1)}
\eea

\subsection{Case (II) - Longitudinal theory} For case (II) we obtain 
the conditions
\bea{ll}
a_1=c,~
&a_2=0\\
a_3=\frac{s(s+1)}{2s+1}c, ~
&a_4=-\frac{s^2}{2s+1}c\\
\eea
and we have to introduce a fermionic compensator $\chi_{\a(s-1)
\ad(s-2)}$ which transforms like 
\bea{ll}
\d_G\chi_{\a(s-1)\ad(s-2)}=&\Dd^{\ad_{s-1}}\D^{\a_s}L_{\a(s)\ad(
s-1)}+\frac{s-1}{s}\D^{\a_s}\Dd^{\ad_{s-1}}L_{a(s)\ad(s-1)}\\
&+\Dd_{\ad_{s-2}}J_{\a(s-1)\ad(s-3)}
\eea
and couples with the term $\Dd^{\bd}\D^{\g}\Dd^{\gd}H_{\g\a(
s-1)\bd\gd\ad(s-2)}$

So in order to achieve invariance we add to the action two new 
pieces, the coupling term of $H$ with $\chi$ and the kinetic 
energy terms for $\chi$. The full action takes the form
\bea{ll}
S=\int d^8z&\ c~ H^{\a(s)\ad(s)}\D^{\g}\Dd^2\D_{\g}H_{\a(s)
\ad(s)}\\
&+\frac{s(s+1)}{2s+1}c~ H^{\a(s)\ad(s)}\D_{\a_s}\Dd^2\D^{\g}
H_{\g\a(s-1)\ad(s)}+c.c.\\
&-\frac{s^2}{2s+1}c~ H^{\a(s)\ad(s)}\D_{\a_s}\Dd_{\ad_s}\D^{\g}
\Dd^{\gd}H_{\g\a(s-1)\gd\ad(s-1)}+c.c.\\
&-\frac{2s^2}{2s+1}c~H^{\a(s)\ad(s)}\Dd_{\ad_s}\D_{\a_s}
\Dd_{\ad_{s-1}}\chi_{\a(s-1)\ad(s-2)}+c.c.\\
&+b_1 \chi^{\a(s-1)\ad(s-2)}\D^2\chi_{\a(s-1)\ad(s-2)}+c.c.
\IEEEyesnumber\\
&+b_2 \chi^{\a(s-1)\ad(s-2)}\Dd^2\chi_{\a(s-1)\ad(s-2)}+c.c.\\
&+b_3 \chi^{\a(s-1)\ad(s-2)}\Dd^{\ad_{s-1}}\D_{\a_{s-1}}\bar{
\chi}_{\a(s-2)\ad(s-1)}\\
&+b_4 \chi^{\a(s-1)\ad(s-2)}\D_{\a_{s-1}}\Dd^{\ad_{s-1}}\bar{
\chi}_{\a(s-2)\ad(s-1)}\\
\eea
and it has to be invariant under
\bea{ll}
\delta_G H_{\a(s)\ad(s)}&=\frac{1}{s!}\D_{(\a_{s}}\bar{L}_{\a(
s-1))\ad(s)}-\frac{1}{s!}\Dd_{(\ad_s}L_{\a(s)\ad(s-1))}
\IEEEyessubnumber\\
\d_G\chi_{\a(s-1)\ad(s-2)}=&\Dd^{\ad_{s-1}}\D^{\a_s}
L_{\a(s)\ad(s-1)}+\frac{s-1}{s}\D^{\a_s}\Dd^{\ad_{s-1}}
L_{a(s)\ad(s-1)}\IEEEyessubnumber\\
&+\Dd_{\ad_{s-2}}J_{\a(s-1)\ad(s-3)}
\eea

The equations of motion of the superfields are
\bea{l}
T_{\a(s)\ad(s)}=\frac{\delta S}{\delta H^{\a(s)\ad(s)}},~G_{\a(s-1)
\ad(s-2)}=\frac{\delta S}{\delta \chi^{\a(s-1)\ad(s-2)}}
\IEEEyesnumber
\eea
and satisfy the Bianchi Identities
\bea{l}
\Dd^{\ad_s}T_{\a(s)\ad(s)}+\frac{1}{s!(s-1)!}\D_{(\a_s}\Dd_{(\ad_{
s-1}} G_{\a(s-1)))\ad(s-2))}\\
~~~~~~~~~~~~~~~+\left[\frac{s-1}{s}\right]\frac{1}{s!(s-1)!}\Dd_{(
\ad_{s-1}}\D_{(\a_s} G_{\a(s-1))\ad(s-2))}=0\IEEEyessubnumber\\
\Dd^{\ad_{s-2}}G_{a(s-1)\ad(s-2)}=0~~~\leadsto\Dd^2G_{a(s-1)
\ad(s-2)}=0\IEEEyessubnumber
\eea
which fix all free coefficients to the following values:
\bea{ll}
b_1=0, ~~ & b_2=\frac{s^2(s+1)}{(2s+1)(s-1)}c\\
b_4=0, ~~ & b_3=\frac{2s^2}{2s+1}c\\
\eea
The superspace action takes the final form
\bea{ll}
\label{S.A.II}
S=\int d^8z&\ c~ H^{\a(s)\ad(s)}\D^{\g}\Dd^2\D_{\g}H_{\a(s)
\ad(s)}\\
&+\left[\frac{s(s+1)}{2s+1}\right]c~ H^{\a(s)\ad(s)}\D_{\a_s}
\Dd^2\D^{\g}H_{\g\a(s-1)\ad(s)}+c.c.\\
&-\left[\frac{s^2}{2s+1}\right]c~ H^{\a(s)\ad(s)}\D_{\a_s}
\Dd_{\ad_s}\D^{\g}\Dd^{\gd}H_{\g\a(s-1)\gd\ad(s-1)}+c.c.\\
&-\left[\frac{2s^2}{2s+1}\right]c~H^{\a(s)\ad(s)}\Dd_{\ad_s}
\D_{\a_s}\Dd_{\ad_{s-1}}\chi_{\a(s-1)\ad(s-2)}+c.c.
\IEEEyesnumber\\
&+\left[\frac{s^2(s+1)}{(2s+1)(s-1)}\right]c~\chi^{\a(s-1)\ad(
s-2)}\Dd^2\chi_{\a(s-1)\ad(s-2)}+c.c.\\
&+\left[\frac{2s^2}{2s+1}\right]c~\chi^{\a(s-1)\ad(s-2)}
\Dd^{\ad_{s-1}}\D_{\a_{s-1}}\bar{\chi}_{\a(s-2)\ad(s-1)}\\
\eea
and the equations of motion are
\bea{ll}
\label{E.Q.II}
T_{\a(s)\ad(s)}&=2c\D^\g\Dd^2\D_\g H_{\a(s)\ad(s)}\\
&~~+\frac{2c}{s!}\left[\frac{s(s+1)}{2s+1}\right]\D_{(\a_s}
\Dd^2\D^\g H_{\g\a(s-1)\ad(s)}\\
&~~+\frac{2c}{s!}\left[\frac{s(s+1)}{2s+1}\right]\Dd_{(\ad_s
}\D^2\Dd^\g H_{\a(s)\gd\ad(s-1))}\\
&~~-\frac{2c}{s!s!}\left[\frac{s^2}{2s+1}\right]\D_{(\a_s}
\Dd_{(\ad_s}\D^\g\Dd^\gd H_{\g\a(s-1))\gd\ad(s-1))}\\
&~~-\frac{2c}{s!s!}\left[\frac{s^2}{2s+1}\right]\Dd_{(\ad_s}
\D_{(\a_s}\Dd^\gd\D^\g H_{\g\a(s-1))\gd\ad(s-1))}
\IEEEyesnumber\\
&~~-\frac{2c}{s!s!}\left[\frac{s^2}{2s+1}\right]\Dd_{(\ad_s}
\D_{(\a_s}\Dd_{\ad_{s-1}}\chi_{\a(s-1))\ad(s-2))}\\
&~~-\frac{2c}{s!s!}\left[\frac{s^2}{2s+1}\right]\D_{(\a_s}
\Dd_{(\ad_s}\D_{\a_{s-1}}\bar{\chi}_{\a(s-2))\ad(s-1))}\\
&{}\\
G_{\a(s-1)\ad(s-2)}&=2c\left[\frac{s^2}{2s+1}\right]
\Dd^{\ad_{s-1}}\D^{\a_s}\Dd^{\ad_s}H_{\a(s)\ad(s)}\\
&~~+2c\left[\frac{s^2(s+1)}{(2s+1)(s-1)}\right]\Dd^2
\chi_{\a(s-1)\ad(s-2)}\IEEEyesnumber\\
&~~+\frac{2c}{(s-1)!}\left[\frac{s^2}{2s+1}\right]\Dd^{
\ad_{s-1}}\D_{(\a_{s-1}}\bar{\chi}_{\a(s-2))\ad(s-1)}
\eea

Using the equations of motion we can now prove that a 
chiral superfield $F_{\a(2s+1)}$ exist and satisfies the 
following identity
\bea{ll}
\D^{\a_{2s+1}}F_{\a(2s+1)}=&\frac{1}{2c}\frac{1}{(2s)!}
\pa_{(\a_{2s}}{}^{\ad_{s}}\ldots\pa_{\a_{s+1}}{}^{\ad_1}
T_{\a(s))\ad(s)}\IEEEyesnumber
\eea
where
\bea{l}
F_{\a(2s+1)}=\frac{1}{(2s+1)!}\Dd^2\D_{(\a_{2s+1}}
\pa_{\a_{2s}}{}^{\ad_{s}}\ldots\pa_{\a_{s+1}}{}^{\ad_{
1}}H_{\a(s))\ad(s)}
\eea
and that proves that in the on-shell theory where $T_{\a(
s)\ad(s-1)}$ = $G_{\a(s)\ad(s-1)}$ = 0 we obtain the desired 
constraints to describe a super-helicity $Y=s+1/2$ system

Unlike the previous theories of half-integer and integer 
super-helicity, we can not perform any local redefinitions 
of the superfields because of the difference in their index 
structure. So the above action is unique.

\section{Projection and Components}
\label{Projection}
~~ The superspace actions derived above in terms of unconstrained 
objects will be the starting point for our component discussion. We will 
use the method described in \cite{IntSpin} to derive the field structure 
of the theory, the component action and their SUSY-transformations laws. 
\subsection{Component structure for Transverse theories (I)}
~~The two superfields $T_{\a(s)\ad(s)}, G_{\a(s)\ad(s-1)}$ in (\ref{E.Q.I}) 
have mass dimensionality $\left[T_{\a(s)\ad(s)}\right]=2$,\\
$\left[G_{\a(s)\ad(s-1)}\right]=3/2$
and satisfy the Bianchi identities and their consequences:
\bea{ll}
\Dd^{\ad_s}T_{\a(s)\ad(s)}-\Dd^2 G_{\a(s))\ad(s-1)}=0 \leadsto\Dd^2
T_{\a(s)\ad(s)}=0\IEEEyessubnumber\\
~~~~~~~~~~~~~~~~~~~~~~~~~~~~~~~~~~~~~~~~~~~~~~~~\D^2
T_{\a(s)\ad(s)}=0~~ \text{reality}\\
\frac{1}{(s+1)!}\D_{(\a_{s+1}}G_{a(s))\ad(s-1)}=0 \leadsto\D^2G_{\a(
s)\ad(s-1)}=0\IEEEyessubnumber
\eea
These identities constrained must of the components of superfields 
$T$ and $G$ and only few of them remain to play the role of off-shell 
auxiliary components. So just by looking at them we immediately see 
the structure of auxiliary fields:
\bea{l}
\Dd^{\ad_{s-1}}G_{\a(s))\ad(s-1)}|,~\Dd_{(\ad_s}G_{\a(s)\ad(s-1))}|,~
T_{\a(s)\ad(s)}|,~\D^{\a_s}G_{\a(s)\ad(s-1)}|~\text{for bosons}\\
G_{\a(s)\ad(s-1)}|,~\D_{(\a_s}\Dd^{\ad_s}\bar{G}_{\a(s-1))\ad(s)}|~~
\text{for fermions}
\eea

The next step is to express the action in terms of $T$ and $G$
\bea{ll}
S=\int d^8z &\left\{~~\frac{1}{2}H^{\a(s)\ad(s)}T_{\a(s)\ad(s)}\right.
\IEEEyesnumber\\
&\left.~+\frac{1}{2}\chi^{\a(s)\ad(s-1)}G_{\a(s)\ad(s-1)}+c.c.\right\}\\
=\int d^4x &\frac{1}{2}\Dd^2\D^2\left(H^{\a(s)\ad(s)}T_{\a(s)\ad(s)}
\right)\\
&+\frac{1}{2}\Dd^2\D^2\left(\chi^{\a(s)\ad(s-1)}G_{\a(s)\ad(s-1)}
\right)+c.c.
\eea
and then to distribute the covariant derivatives.
\subsubsection{Fermions}
~~ After the distribution of $\D$'s and the usage of Bianchi identities 
we derive for the fermionic Lagrangian:
\bea{lll}
\mathcal{L}_F&=&\frac{1}{2}\D^2\Dd^{\ad_{s+1}}H^{\a(s)\ad(s)}|\frac{
1}{(s+1)!}\Dd_{(\ad_{s+1}}T_{\a(s)\ad(s))}|\\
&+&\frac{1}{2}\left(-\frac{s}{s+1}\D^2\Dd_{\gd}H^{\a(s)\gd\ad(s-1)}+
\D^2\chi^{\a(s)\ad(s-1)}\right)|\Dd^{\ad_s}T_{\a(s)\ad(s)}|\\
&+&\frac{1}{2}\frac{s}{s+1}\Dd^{\ad_s}\D_{\g}\chi^{\g\a(s-1)\ad(s-1)}|\frac{
1}{s!}\Dd_{(\ad_{s}}\D^{\a_s}G_{\a(s)\ad(s-1))}|\\
&-&\frac{1}{2}\frac{s-1}{s+1}\Dd_{\gd}\D_{\g}\chi^{\g\a(s-1)\gd\ad(s-2)}|
\Dd^{\ad_{s-1}}\D^{\a_s}G_{\a(s)\ad(s-1)}|\IEEEyesnumber\\
&+&\frac{1}{2}\Dd^2\D^2\chi^{\a(s)\ad(s-1)}|G_{\a(s)\ad(s-1)}|\\
&+&c.c.
\eea
$T$ and $G$ satisfy a few more identities:
\bea{ll}
&\frac{1}{(s+1)!}\Dd_{(\ad_{s+1}}T_{\a(s)\ad(s))}=\\
&=\frac{2ic}{(s+1)!^2}\pa^{\a_{s+1}}{}_{(\ad_{s+1}}\Dd^2\D_{(\a_{s+1}}
H_{\a(s))\ad(s))}\IEEEyesnumber\\
&~~-\frac{2ic}{(s+1)!s!}\frac{s}{s+1}\pa_{(\a_{s}(\ad_{s+1}}\left[\Dd^2\D^{
\g}H_{\g\a(s-1))\ad(s))}-\frac{s+1}{s}\Dd^2\bar{\chi}_{\a(s-1)\ad(s)}\right]
\eea

%%%%
%%%%
\bea{ll}
\Dd^{\ad_{s-1}}\D^{\a_s}G_{\a(s)\ad(s-1)}&=i\frac{s+1}{s}\pa^{\a_s\ad_{
s-1}}\left[\vphantom{\frac12}G_{\a(s)\ad(s-1))}+2c\D^2\Dd^{\ad_s}H_{
\a(s)\ad(s)}\right.\\
&~~~~~~~~~~~~~~~~~~~~~~~~~~\left.+2c\frac{s+1}{s}\D^2\chi_{\a(s)
\ad(s-1)}\right]\IEEEyesnumber\\
&~~-\frac{2ic}{(s-1)!}\frac{s^2-1}{s^2}\pa_{(\a_{s-1}}{}^{\ad_{s-1}}
\D^{\g}\Dd^{\gd}\bar{\chi}_{\g\a(s-2)\gd\ad(s-1)}\\
\eea

%%%%
%%%%
\bea{ll}
\Dd^{\ad_{s}}T_{\a(s)\ad(s)}&=\frac{2ic}{(s+1)!}\pa^{\a_{s+1}\ad_s}
\Dd^2\D_{(\a_{s+1}}H_{\a(s))\ad(s)}\\
&+\frac{2ic}{s!}\frac{2s+1}{s(s+1)}\pa_{(\a_s}{}^{\ad_s}\left[\Dd^2
\D^{\g}H_{\g\a(s-1))\ad(s)}-\frac{s+1}{s}\Dd^2\bar{\chi}_{\a(s-1))\ad(
s)}\right]\IEEEyesnumber\\
&+\frac{2ic}{s!(s-1)!}\frac{s^2-1}{s^2}\pa_{(\a_s(\ad_{s-1}}\Dd^{\gd}
\D^{\g}\chi_{\g\a(s-1)\gd\ad(s-2))}\\
&+\frac{1}{s!}\D_{(\a_s}\Dd^{\ad_s}\bar{G}_{\a(s-1))\ad(s)}\\
&-\frac{i}{s!}\frac{s+1}{s}\pa_{(\a_s}{}^{\ad_s}\bar{G}_{\a(s-1))\ad(s)}
\eea

We observe that in all the above expressions and in the fermionic 
Lagrangian there are some specific combinations that appear 
repeatedly.  So let us define
\bea{l}
\frac{1}{(s+1)!}\Dd^2\D_{(\a_{s+1}}H_{\a(s))\ad(s)}|\equiv N_1\psi_{\a(s+
1)\ad(s)}\\
\left\{\D^2\Dd^{\ad_s}H_{\a(s)\ad(s)}+\frac{s+1}{s}\D^2\chi_{\a(s)\ad(s-1
)}\right\}|\equiv N_2\psi_{\a(s)\ad(s-1)}\IEEEyesnumber\\
\Dd^{\ad_{s-1}}\D^{\a_s}\chi_{\a(s)\ad(s-1)}|\equiv N_3\psi_{\a(s-1)\ad(
s-2)}\\
\eea
where $N_1,~N_2,~N_3$ are normalization constants to be fixed later.
%%%%
%%%%
Putting everything together we have for the Lagrangian
\bea{ll}
\mathcal{L}_F=&G^{\a(s)\ad(s-1)}|\left(-\frac{1}{2c}\frac{s}{s+1}\frac{1
}{s!}\D_{(\a_s}\Dd^{\ad_s}\bar{G}_{\a(s-1))\ad(s)}\right.\\
&~~~~~~~~~~~~~~~~~~~~\left.+\frac{i}{4c}\frac{1}{s!}\pa_{(\a_s}{}^{
\ad_s}\bar{G}_{\a(s-1))\ad(s)}\right)|+c.c.\\
&+2ic|N_1|^2\bar{\psi}^{\a(s)\ad(s+1)}\pa^{\a_{s+1}}{}_{\ad_{s+1}}
\psi_{\a(s+1)\ad(s)}\\
&-2ic\frac{s}{s+1}N_1N_2\psi^{\a(s+1)\ad(s)}\pa_{\a_{s+1}\ad_s}
\psi_{\a(s)\ad(s-1)}+c.c.\IEEEyesnumber\\
&-2ic\frac{2s+1}{(s+1)^2}|N_2|^2\bar{\psi}^{\a(s-1)\ad(s)}\pa^{\a_{s
}}{}_{\ad_{s}}\psi_{\a(s)\ad(s-1)}\\
&+2ic\frac{s-1}{s}N_2N_3\psi^{\a(s)\ad(s-1)}\pa_{\a_{s}\ad_{s-1}}
\psi_{\a(s-1)\ad(s-2)}+c.c.\\
&-2ic\left(\frac{s-1}{s}\right)^2|N_3|^2\bar{\psi}^{\a(s-2)\ad(s-1)}
\pa^{\a_{s-1}}{}_{\ad_{s-1}}\psi_{\a(s-1)\ad(s-2)}\\
\eea
The first term in the Lagrangian is the algebraic kinetic energy term 
of two auxiliary fields and the rest of the terms are exactly the structure 
of a theory that describes helicity $h=s+1/2$~\cite{BK}\footnote{Using 
the conventions of \cite{GGRS}}. To have an exact match we choose 
coefficients
\bea{ll}
c=1~,~& N_2=-\frac{1}{\sqrt{2}}\\
N_1=\frac{1}{\sqrt{2}}~,~& N_3=-\frac{1}{\sqrt{2}}\frac{s}{s-1}
\eea
So the fields that appear in the fermionic action are defined as:
\bea{l}
\rho_{\a(s)\ad(s-1)}\equiv G_{\a(s)\ad(s-1)}|\\
\b_{\a(s)\ad(s-1)}\equiv-\frac{1}{2s!}\left\{\frac{s}{s+1}\D_{(\a_s}\Dd^{
\ad_s}\bar{G}_{\a(s-1))\ad(s)}-\frac{i}{2}\pa_{(\a_s}{}^{\ad_s}\bar{G}_{
\a(s-1))\ad(s)}\right\}|\\
\psi_{a(s+1)\ad(s)}\equiv\frac{\sqrt{2}}{(s+1)!}\Dd^2\D_{(\a_{s+1}}
H_{\a(s))\ad(s)}|\IEEEyesnumber\\
\psi_{\a(s)\ad(s-1)}\equiv-\sqrt{2}\left\{\D^2\Dd^{\ad_s}H_{\a(s)\ad(
s)}+\frac{s+1}{s}\D^2\chi_{\a(s)\ad(s-1)}\right\}|\\
\psi_{\a(s-1)\ad(s-2)}\equiv-\sqrt{2}\frac{(s-1)}{s}\Dd^{\ad_{s-1}}
\D^{\a_s}\chi_{\a(s)\ad(s-1)}|
\eea
The Lagrangian is
\bea{ll}
\mathcal{L}_F=&\rho^{\a(s)\ad(s-1)}\beta_{\a(s)\ad(s-1)}+c.c.\\
&+i~\bar{\psi}^{\a(s)\ad(s+1)}\pa^{\a_{s+1}}{}_{\ad_{s+1}}\psi_{\a(
s+1)\ad(s)}\\
&+i\left[\frac{s}{s+1}\right]~\psi^{\a(s+1)\ad(s)}\pa_{\a_{s+1}\ad_s}
\psi_{\a(s)\ad(s-1)}+c.c.\\
&-i\left[\frac{2s+1}{(s+1)^2}\right]~\bar{\psi}^{\a(s-1)\ad(s)}\pa^{\a_{
s}}{}_{\ad_{s}}\psi_{\a(s)\ad(s-1)}\IEEEyesnumber\\
&+i~\psi^{\a(s)\ad(s-1)}\pa_{\a_{s}\ad_{s-1}}\psi_{\a(s-1)\ad(s-2)}+c.c.\\
&-i~\bar{\psi}^{\a(s-2)\ad(s-1)}\pa^{\a_{s-1}}{}_{\ad_{s-1}}\psi_{\a(s-1
)\ad(s-2)}\\
\eea
and the gauge transformations of the fields are
\bea{ll}
\d_G\rho_{\a(s)\ad(s-1)}=0~,~&\d_G\psi_{\a(s+1)\ad(s)}=\frac{1}{s!(
s+1)!}\pa_{(\a_{s+1}(\ad_s}\xi_{\a(s))\ad(s-1))}\\
\d_G\b_{\a(s)\ad(s-1)}=0~,~&\d_G\psi_{\a(s)\ad(s-1)}=-\frac{1}{s!}
\pa_{(\a_{s}}{}^{\ad_s}\bar{\xi}_{\a(s-1))\ad(s)}\IEEEyesnumber\\
&\d_G\psi_{\a(s-1)\ad(s-2)}=\frac{s-1}{s}\pa^{\a_s\ad_{s-1}}\xi_{\a(
s)\ad(s-1)}\\
& \text{with } \xi_{\a(s)\ad(s-1)}=-i\sqrt{2}~\Dd^2L_{\a(s)\ad(s-1)}|
\eea
\subsubsection{Bosons}
For the bosonic action we follow exactly the same procedure. The 
fields that appear in the action are defined as:
\bea{l}
U_{\a(s)\ad(s-2)}\equiv\Dd^{\ad_{s-1}}G_{\a(s))\ad(s-1)}|\\
u_{\a(s)\ad(s)}\equiv\frac{1}{2s!}\left\{\D_{(\a_s}\bar{G}_{\a(s-1))
\ad(s)}-\Dd_{(\ad_s}G_{\a(s)\ad(s-1))}\right\}|\\
v_{\a(s)\ad(s)}\equiv -\frac{i}{2s!}\left\{\D_{(\a_s}\bar{G}_{\a(s-1))
\ad(s)}+\Dd_{(\ad_s}G_{\a(s)\ad(s-1))}\right\}|\\
A_{\a(s)\ad(s)}\equiv T_{\a(s)\ad(s)}|+\frac{s}{2s+1}\frac{1}{s!}
\left(\D_{(\a_s}\bar{G}_{\a(s-1))\ad(s)}-\Dd_{(\ad_s}G_{\a(s)\ad(s-
1))}\right)|\\
S_{\a(s-1)\ad(s-1)}\equiv\frac{1}{2}\left\{\D^{\a_s}G_{\a(s)\ad(s-1)}
+\Dd^{\ad_s}\bar{G}_{\a(s)\ad(s-1)}\right\}|\IEEEyesnumber\\
P_{\a(s-1)\ad(s-1)}\equiv -\frac{i}{2}\left\{\D^{\a_s}G_{\a(s)\ad(s-1
)}-\Dd^{\ad_s}\bar{G}_{\a(s)\ad(s-1)}\right\}|\\
h_{\a(s+1)\ad(s+1)}\equiv\frac{1}{2}\frac{1}{(s+1)!^2}\left[\D_{(\a_{
s+1}},\D_{(\ad_{s+1}}\right]H_{\a(s))\ad(s))}|\\
h_{\a(s-1)\ad(s-1)}\equiv\frac{1}{2}\frac{s}{(s+1)^2}\left[\D^{\a_{s}},
\Dd^{\ad_{s}}\right]H_{\a(s)\ad(s)}|\\
~~~~~~~~~~~~~~~~~~~~+\frac{1}{s+1}\left(\D^{\a_s}\chi_{\a(s)
\ad(s-1)}+\Dd^{\ad_{s}}\bar{\chi}_{\a(s-1)\ad(s)}\right)|
\eea
the gauge transformations are
\bea{l}
\d_G U_{\a(s)\ad(s-2)}=0,~~\d_G A_{\a(s)\ad(s)}=0\\
\d_G u_{\a(s)\ad(s)}=0,~~~~~\d_G S_{\a(s-1)\ad(s-1)}=0
\IEEEyesnumber\\
\d_G v_{\a(s)\ad(s)}=0,~~~~~\d_G P_{\a(s-1)\ad(s-1)}=0\\
\d_G h_{\a(s+1)\ad(s+1)}=\frac{1}{(s+1)!^2}\pa_{(\a_{s+1}(\ad_{
s+1}}\zeta_{\a(s))\ad(s))}\\
\d_G h_{\a(s-1)\ad(s-1)}=\frac{s}{(s+1)^2}\pa^{\a_{s}\ad_{s}}
\zeta_{\a(s)\ad(s)}\\
\eea
where
\bea{l}
\zeta_{\a(s)\ad(s)}=\frac{i}{2s!}\left(\D_{(\a_s}\bar{L}_{\a(s-1))\ad(
s)}+\Dd_{(\ad_s}L_{\a(s)\ad(s-1))}\right)|
\eea
and the Lagrangian
\bea{ll}
\mathcal{L}_B=&\frac{1}{4}\left[\frac{s-1}{s+1}\right]~U^{\a(s)\ad(
s-2)}U_{\a(s)\ad(s-2)}+c.c.\\
&+\frac{1}{2}\left[\frac{s}{2s+1}\right]~u^{\a(s)\ad(s)}u_{\a(s)\ad(s)}\\
&-\left[\frac{s}{2}\right]~v^{\a(s)\ad(s)}v_{\a(s)\ad(s)}\\
&+\frac{1}{8}\left[\frac{2s+1}{s+1}\right]~A^{\a(s)\ad(s)}A_{\a(s)\ad(
s)}\\
&-\frac{1}{2}\left[\frac{s^2}{(s+1)^2}\right]~S^{\a(s-1)\ad(s-1)}S_{
\a(s-1)\ad(s-1)}\\
&-\frac{1}{2}\left[\frac{s^2}{(s+1)^2}\right]~P^{\a(s-1)\ad(s-1)}P_{
\a(s-1)\ad(s-1)}\\
&+~h^{\a(s+1)\ad(s+1)}\Box h_{\a(s+1)\ad(s+1)}\IEEEyesnumber\\
&-\left[\frac{s+1}{2}\right]~h^{\a(s+1)\ad(s+1)}\pa_{\a_{s+1}\ad_{s+1
}}\pa^{\g\gd}h_{\g\a(s)\gd\ad(s)}\\
&+\left[s(s+1)\right]~h^{\a(s+1)\ad(s+1)}\pa_{\a_{s+1}\ad_{s+1}}\pa_{
\a_{s}\ad_{s}}h_{\a(s-1)\ad(s-1)}\\
&-\left[(s+1)(2s+1)\right]~h^{\a(s-1)\ad(s-1)}\Box h_{\a(s-1)\ad(s-1)}\\
&-\left[\frac{(s+1)(s-1)^2}{2}\right]~h^{\a(s-1)\ad(s-1)}\pa_{\a_{s-1}
\ad_{s-1}}\pa^{\g\gd}h_{\g\a(s-2)\gd\ad(s-2)}
\eea
gives rise to the theory of  helicity $h=s+1$ as expected.
\subsubsection{Off-shell degrees of freedom}
Let us count the bosonic degrees of freedom:
\begin{center}
\begin{tabular}{|c|c|c|c|}
\hline 
\emph{fields} & \emph{d.o.f} & \emph{redundancy} & \emph{net} \\ 
\hline 
$h_{\a(s+1)\ad(s+1)}$ & $(s+2)^2$ & \multirow{2}{*}{$(s+1)^2$} & 
\multirow{2}{*}{$s^2+2s+3$} \\ 
\cline{1-2}
$h_{\a(s-1)\ad(s-1)}$ & $s^2$ & & \\ 
\hline 
$u_{\a(s)\ad(s)}$ & $(s+1)^2$ & 0 & $(s+1)^2$  \\ 
\hline 
$v_{\a(s)\ad(s)}$ & $(s+1)^2$ & 0 & $(s+1)^2$ \\ 
\hline 
$A_{\a(s)\ad(s)}$ & $(s+1)^2$ & 0 & $(s+1)^2$ \\ 
\hline 
$U_{\a(s)\ad(s-2)}$ & $2(s+1)(s-1)$ & 0 & $2(s+1)(s-1)$ \\ 
\hline 
$S_{\a(s-1)\ad(s-1)}$ & $s^2$ & 0 & $s^2$ \\ 
\hline 
$P_{\a(s-1)\ad(s-1)}$ & $s^2$ & 0 & $s^2$ \\ 
\hline\hline 
\multicolumn{2}{c|}{} & \emph{Total} & $8s^2+8s+4$ \\ 
\cline{3-4}
\end{tabular}
\end{center}
and the same counting for the fermionic degrees of freedom:
\begin{center}
\begin{tabular}{|c|c|c|c|}
\hline 
\emph{fields} & \emph{d.o.f} & \emph{redundancy} & \emph{net} \\ 
\hline 
$\psi_{\a(s+1)\ad(s)}$ & $2(s+2)(s+1)$ & \multirow{3}{*}{$2(s+1)
s$} & \multirow{3}{*}{$4s^2+4s+4$} \\ 
\cline{1-2}
$\psi_{\a(s)\ad(s-1)}$ & $2(s+1)s$ & & \\ 
\cline{1-2}
$\psi_{\a(s-1)\ad(s-2)}$ & $2s(s-1)$ & & \\ 
\hline 
$\rho_{\a(s)\ad(s-1)}$ & $2(s+1)s$ & 0 & $2(s+1)s$ \\ 
\hline 
$\b_{\a(s)\ad(s-1)}$ & $2(s+1)s$ & 0 & $2(s+1)s$ \\ 
\hline\hline 
\multicolumn{2}{c|}{} & \emph{Total} & $8s^2+8s+4$ \\ 
\cline{3-4}
\end{tabular}
\end{center}

\subsubsection{SUSY-transformation laws}
The last thing left in order to complete the component picture, 
is to find the SUSY-transformation laws of the fields. They can 
be calculated by the action of the SUSY-generators on the 
specific component. In terms of the covariant derivatives we obtain
\be
\d_S\text{Component}=-\left(\e^{\b}\D_{\b}+\ed^{\bd}\Dd_{\bd}
\right)\text{Component}|\nonumber
\ee
For the dynamical fields (they have non-zero gauge transformation) 
the redundancy allows us to ignore all the terms that have the same 
structure as their gauge transformation law because of the identification
\bea{c}
\d_S\{\text{Dynamical field}\}\sim\d_S\{\text{Dynamical field}\}+\pa
\left(\d_S\mathcal{\zeta}\right)
\eea

With all that in mind, the transformation of the fermionic fields are:
\bea{ll}
\d_S\rho_{\a(s)\ad(s-1)}=&\frac{s}{s+1}\frac{1}{s!}\e_{(\a_s}\left[S_{
\a(s-1))\ad(s-1)}+iP_{\a(s-1))\ad(s-1)}\right]\\
&+\ed^{\ad_s}\left[u_{\a(s)\ad(s)}-iv_{\a(s)\ad(s)}\right]\IEEEyesnumber\\
&+\frac{s-1}{s}\frac{1}{(s-1)!}\ed_{(\ad_{s-1}}U_{\a(s)\ad(s-2))}
\eea
\bea{ll}
\d_S\b_{\a(s)\ad(s-1)}=&\frac{s}{s+1}\frac{1}{s!}\e_{(\a_s}\pa^{\g\gd}
A_{\g\a(s-1))\gd\ad(s-1)}\\
&\frac{2s}{(s+1)^2}\frac{i}{s!}\e_{(\a_s}\pa^{\g\gd}u_{\g\a(s-1)\gd\ad(
s-1)}\\
&\frac{2s}{s!}\e_{(\a_s}\pa^{\g\gd}v_{\g\a(s-1)\gd\ad(s-1)}\\
&-\frac{i}{s!}\e^{\g}\pa_{(\a_s}{}^{\gd}u_{\g\a(s-1))\gd\ad(s-1)}\\
&+\frac{1}{s!}\e^{\g}\pa_{(\a_s}{}^{\gd}v_{\g\a(s-1))\gd\ad(s-1)}\\
&+\frac{i}{s!}\frac{2s-1}{s+1}\ed^{\gd}\pa_{(\a_s\gd}\left[S_{\a(s-1))
\ad(s-1)}-iP_{\a(s-1))\ad(s-1)}\right]\\
&+\frac{i}{s!(s-1)!}\frac{s-1}{s+1}\ed_{(\ad_{s-1}}\pa_{(\a_s}{}^{\gd}
\left[S_{\a(s-1))\gd\ad(s-2))}\right.\IEEEyesnumber\\
&~~~~~~~~~~~~~~~~~~~~~~~~~~~~~~~~~~~~~~\left.-iP_{\a(s-1))
\gd\ad(s-2))}\right]\\
&+\frac{2i}{s!(s-1)!}\frac{s-1}{s+1}\e_{(\a_s}\pa^{\g}{}_{(\ad_{s-1}}
U_{\g\a(s-1))\ad(s-2))}\\
&+\frac{i}{s!}\frac{s-1}{s}\e_{(\a_s}\pa_{\a_{s-1}}{}^{\gd}\bar{U}_{
\a(s-2))\gd\ad(s-1)}\\
&+\frac{2s}{s!}\e_{(\a_s}\pa^{\g\gd}\pa^{\b\bd}h_{\b\g\a(s-1))\bd
\gd\ad(s-1)}\\
&-\frac{2(s-1)^2}{s!(s-1)!}\e_{(\a_s}\pa_{\a_{s-1}(\ad_{s-1}}\pa^{
\g\gd}h_{\g\a(s-2))\gd\ad(s-2))}
\eea
\bea{ll}
\d_S\psi_{\a(s+1)\ad(s)}=&\frac{\sqrt{2}i}{(s+1)!}\e^{\g}\pa_{(\a_{
s+1}}{}^{\gd}h_{\g\a(s))\gd\ad(s)}\\
&-\frac{i}{\sqrt{2}(s+1)!}\e_{(\a_{s+1}}\pa^{\g\gd}h_{\g\a(s))\gd\ad(
s)}\IEEEyesnumber\\
&+\frac{1}{2\sqrt{2}}\frac{2s+1}{s+1}\frac{1}{(s+1)!}\e_{(\a_{s+1}}
A_{\a(s))\ad(s)}
\eea
\bea{ll}
\d_S\psi_{\a(s)\ad(s-1)}=&-\frac{1}{2\sqrt{2}}\frac{s}{s+1}\ed^{
\ad_s}A_{\a(s)\ad(s)}\\
&+\frac{1}{\sqrt{2}}\frac{s+1}{2s+1}\ed^{\ad_s}u_{\a(s)\ad(s)}\\
&-i\frac{s+1}{\sqrt{2}}\ed^{\ad_s}v_{\a(s)\ad(s)}\\
&+\frac{1}{\sqrt{2}}\frac{s-1}{s!}\ed_{(\ad_{s-1}}U_{\a(s)\ad(s-2))}
\IEEEyesnumber\\
&-\frac{is}{\sqrt{2}}\ed^{\ad_s}\pa^{\g\gd}h_{\g\a(s)\gd\ad(s)}\\
&\frac{i}{s!^2}\frac{s(s+2)}{\sqrt{2}}\ed^{\ad_s}\pa_{(\a_s(\ad_s}
h_{\a(s-1))\ad(s-1))}
\eea
\bea{ll}
\d_S\psi_{\a(s-1)\ad(s-2)}=&-\frac{1}{\sqrt{2}}\frac{s-1}{s+1}
\e^{\a_s}U_{\a(s)\ad(s-2)}\\
&-\frac{1}{\sqrt{2}}\frac{s-1}{s+1}\ed^{\ad_{s-1}}\left[S_{\a(s-1)
\ad(s-1)}-iP_{\a(s-1)\ad(s-1)}\right]\\
&+i\sqrt{2}\frac{s-1}{s!}\e^{\a_s}\pa_{(\a_s}{}^{\ad_{s-1}}h_{\a(s-1))\ad(s-1)}
\eea
and the SUSY-transformation laws for the bosonic fields are:
\bea{ll}
\d_S A_{\a(s)\ad(s)}=&-\frac{i\sqrt{2}}{(s+1)!}\ed^{\ad_{s+1}}\pa^{
\a_{s+1}}{}_{(\ad_{s+1}}\psi_{\a(s+1)\ad(s))}+c.c.\\
&+\frac{i\sqrt{2}}{s!}\frac{s^2}{(2s+1)(s+1)}\ed_{(\ad_s}\pa^{\g\gd}
\psi_{\g\a(s)\gd\ad(s-1))}+c.c.\\
&-\frac{i}{s!(s+1)!}\frac{s}{2s+1}\ed^{\ad_{s+1}}\pa_{(\a_{s}(\ad_{s+1
}}\bar{\rho}_{\a(s-1))\ad(s))}+c.c.\\
&-\frac{i}{s!^2}\frac{s}{(2s+1)(s+1)}\ed_{(\ad_{s}}\pa_{(\a_{s}}{}^{\gd}
\bar{\rho}_{\a(s-1))\gd\ad(s-1))}+c.c.\IEEEyesnumber\\
&+\frac{i\sqrt{2}}{s!(s+1)!}\frac{s}{s+1}\ed^{\ad_{s+1}}\pa_{(\a_s(
\ad_{s+1}}\bar{\psi}_{\a(s-1))\ad(s)}+c.c.\\
&+\frac{i\sqrt{2}}{s!^2}\frac{s}{(s+1)^2}\ed_{(\ad_{s}}\pa_{(\a_s}{}^{
\gd}\bar{\psi}_{\a(s-1))\gd\ad(s-1))}+c.c.\\
&-\frac{i\sqrt{2}}{s!^2}\frac{s}{2s+1}\ed_{(\ad_{s}}\pa_{(\a_s\ad_{s-1
}}\psi_{\a(s-1))\ad(s-2))}+c.c.
\eea
\bea{ll}
\d_S\left(u_{\a(s)\ad(s)}+iv_{\a(s)\ad(s)}\right)=&-\frac{i\sqrt{2}}{s!}
\e_{(\a_s}\pa^{\g\gd}\bar{\psi}_{\g\a(s-1))\gd\ad(s)}\\
&+\frac{i\sqrt{2}}{s!^2}\frac{2s+1}{s(s+1)}\e_{(\a_s}\pa^{\g}{}_{(\ad_s}
\psi_{\g\a(s-1))\ad(s-1))}\\
&+\frac{i\sqrt{2}}{s!^2}\frac{s+1}{s}\e_{(\a_s}\pa_{\a_{s-1}(\ad_s}\bar{
\psi}_{\a(s-2))\ad(s-1))}\\
&-\frac{2}{s!}\frac{s+1}{s}\e_{(\a_s}\bar{\b}_{\a(s-1))\ad(s)}
\IEEEyesnumber\\
&+\frac{2}{s!}\ed_{(\ad_s}\b_{\a(s)\ad(s-1))}\\
&-\frac{i}{s!^2}\frac{s+1}{2s}\e_{(\a_s}\pa^{\g}{}_{\ad_s}\rho_{\g\a(s-1
))\ad(s-1))}\\
&-\frac{i}{s!(s+1)!}\ed^{\ad_{s+1}}\pa_{(\a_s(\ad_{s+1}}\bar{\rho}_{\a(
s-1))\ad(s)}\\
&+\frac{i}{s!^2}\frac{s-1}{2(s+1)}\ed_{(\ad_s}\pa_{(\a_s}{}^{\gd}\bar{
\rho}_{\a(s-1))\gd\ad(s-1))}
\eea
\bea{ll}
\d_S U_{\a(s)\ad(s-2)}=&i\sqrt{2}\ed^{\ad_{s-1}}\pa^{\a_{s+1}\ad_{s}}
\psi_{\a(s+1)\ad(s)}\\
&+\frac{i\sqrt{2}}{s!}\frac{2s+1}{s(s+1)}\ed^{\ad_{s-1}}\pa_{(\a_s}{}^{
\ad_s}\bar{\psi}_{\a(s-1))\ad(s)}\\
&+\frac{i\sqrt{2}}{s!}\e_{(\a_s}\pa^{\g\gd}\psi_{\g\a(s-1))\gd\ad(s-2)}\\
&+\frac{i\sqrt{2}}{s!}\e_{(\a_s}\pa_{\a_{s-1}}{}^{\ad_{s-1}}\bar{\psi}_{
\a(s-2))\ad(s-1)}\\
&-\frac{i\sqrt{2}}{s!(s-1)!}\frac{s+1}{s}\ed^{\ad_{s-1}}\pa_{(\a_s(\ad_{
s-1}}\psi_{\a(s-1))\ad(s-2))}\IEEEyesnumber\\
&-\frac{i}{s!}\frac{1}{s+1}\e_{(\a_s}\pa^{\g\gd}\rho_{\g\a(s-1))\gd\ad(
s-2)}\\
&-\frac{i}{(s+1)!}\e^{\a_{s+1}}\pa_{(\a_{s+1}}{}^{\ad_{s-1}}\rho_{\a(s))
\ad(s-1)}\\
&-\frac{i}{s!}\frac{s+1}{2s}\ed^{\ad_{s-1}}\pa_{(\a_s}{}^{\ad_{s}}\bar{
\rho}_{\a(s-1))\ad(s)}\\
&-2\frac{s+1}{s}\ed^{\ad_{s-1}}\b_{\a(s)\ad(s-1)}
\eea
\bea{ll}
\d_S\left(S_{\a(s-1)\ad(s-1)}\right. &\left.+iP_{\a(s-1)\ad(s-1)}\right)=\\
&=2\frac{s+1}{s}\ed^{\ad_s}\bar{\b}_{\a(s-1)\ad(s)}\\
&~-\frac{i}{s!}\frac{s+1}{2s}\ed^{\ad_s}\pa^{\a_s}{}_{(\ad_s}\rho_{\a(s)
\ad(s-1))}\\
&~+\frac{i}{(s-1)!}\frac{(s-1)(s+1)}{s^2}\ed_{(\ad_{s-1}}\pa^{\g\gd}\rho_{
\g\a(s-1)\gd\ad(s-2)}\IEEEyesnumber\\
&-\frac{i\sqrt{2}}{(s-1)!}\frac{(s-1)(s+1)}{s^2}\ed_{(\ad_{s-1}}\pa^{\g\gd}
\psi_{\g\a(s-1)\gd\ad(s-2))}\\
&-\frac{i\sqrt{2}}{(s-1)!^2}\frac{(s-1)(s+1)}{s^2}\ed_{(\ad_{s-1}}\pa_{(\a_{
s-1}}{}^{\gd}\bar{\psi}_{\a(s-2))\gd\ad(s-2))}
\eea
\bea{ll}
\d_S h_{\a(s+1)\ad(s+1)}=&\frac{1}{\sqrt{2}(s+1)!}\e_{(\a_{s+1}}\bar{\psi
}_{\a(s))\ad(s+1)}+c.c.\IEEEyesnumber
\eea
\bea{ll}
\d_S h_{\a(s-2)\ad(s-2)}=&\frac{1}{\sqrt{2}(s+1)^2}\e^{\a_s}\psi_{\a(s)\ad(
s-1)}+c.c.\\
&-\frac{1}{2(s+1)}\e^{\a_s}\rho_{\a(s)\ad(s-1)}+c.c.\IEEEyesnumber\\
&-\frac{1}{\sqrt{2}(s+1)}\frac{1}{(s-1)!}\ed_{(\ad_{s-1}}\psi_{\a(s-1)\ad(
s-2))}
\eea
\subsection{Component structure for Longitudinal theories (II)}
~~We repeat  the same steps for the second formulation of half-integer 
super-helicity theories.  The superspace action (\ref{S.A.II}) can be 
expressed like
\bea{ll}
S=\int d^8z &\left\{~~\frac{1}{2}H^{\a(s)\ad(s)}T_{\a(s)\ad(s)}\right.
\IEEEyesnumber\\
&\left.~+\frac{1}{2}\chi^{\a(s-1)\ad(s-2)}G_{\a(s-1)\ad(s-2)}+c.c.\right\}\\
=\int d^4x &\frac{1}{2}\D^2\Dd^2\left(H^{\a(s)\ad(s)}T_{\a(s)\ad(s)}\right)\\
&+\frac{1}{2}\D^2\Dd^2\left(\chi^{\a(s-1)\ad(s-2)}G_{\a(s-1)\ad(s-2)}\right)+c.c.
\eea
where $T,~G$ are defined by (\ref{E.Q.II})
\subsubsection{Fermions}
~~ For the fermionic Lagrangian we have
\bea{l}
\mathcal{L}_F=\\
=\frac{1}{2}\frac{1}{(s+1)!}\D^2\Dd^{(\ad_{s+1}}H^{\a(s)\ad(s))}|\frac{1}{(
s+1)!}\Dd_{(\ad_{s+1}}T_{\a(s)\ad(s))}|\\
+\left(\frac{1}{2}\frac{1}{s+1}\D^2\Dd_{\gd}H^{\a(s)\gd\ad(s-1)}
-\frac{1}{2}\frac{1}{s!(s-1)!}\D^{(\a_s}\Dd^{(\ad_{s-1}}\chi^{\a(s-1))\ad(s-2
))}\right.\\
~~~~-\left.\frac{i}{2}\frac{1}{s!}\D^{(\a_s}\pa_{\g\gd}H^{\g\a(s-1))\gd\ad(s-1
)}\right)|\frac{1}{s!(s-1)!}\D_{(\a_s}\Dd_{(\ad_{s-1}}G_{\a(s-1))\ad(s-2))}|\\
+\left(\frac{1}{2}\frac{s-1}{s}\frac{1}{(s-1)!}\D_{\g}\Dd^{(\ad_{s-1}}\chi^{\g
\a(s-2)\ad(s-2))}\right.\IEEEyesnumber\\
~~~~+\left.\frac{i}{2}\frac{s-1}{s}\D_{\b}\pa_{\g\gd}H^{\b\g\a(s-2)\gd\ad(s-
1)}\right)|\frac{1}{(s-1)!}\D^{\a_{s-1}}\Dd_{(\ad_{s-1}}G_{\a(s-1)\ad(s-2))}|\\
+\left(-\frac{i}{2}\frac{s-1}{s+1}\pa_{\a_s\ad_{s-1}}\D^2\Dd_{\ad_s}H_{\a(s)
\ad(s)}\right.\\
~~~~+\left.\frac{1}{2}\D^2\Dd^2\chi^{\a(s-1)\ad(s-2)}\right)|G_{\a(s-1)\ad(
s-2)}|\\
+\frac{1}{2}\Dd^2\chi^{\a(s-1)\ad(s-2)}|\D^2G_{\a(s-1)\ad(s-2)}|\\
+c.c.
\eea

We can prove the following identities for $T$ and $G$:
\bea{l}
\frac{1}{(s+1)!}\Dd_{(\ad_{s+1}}T_{\a(s)\ad(s))}=\\
~~~=\frac{2ic}{(s+1)!}\pa^{\a_{s+1}}{}_{(\ad_{s+1}}\left\{\frac{1}{(s+1)!}
\Dd^2\D_{(\a_{s+1}}H_{\a(s))\ad(s))}\right\}\IEEEyesnumber\\
~~~~-\frac{2ic}{(s+1)!s!}\frac{s^2}{(2s+1)(s+1)}\pa_{(\a_{s}(\ad_{s+1}}
\left\{\vphantom{\frac{1}{2}}\Dd^2\D^{\g}H_{\g\a(s-1))\ad(s))}\right.\\
~~~~~~~~~~~~~~~~~~~~~~~~~~~~~~~~~~~~~~~~~~~~~~~~~+
\frac{i(s+1)}{s!}\Dd_{\ad_s}\pa^{\g\gd}H_{\g\a(s-1))\gd\ad(s-1)))}\\
~~~~~~~~~~~~~~~~~~~~~~~~~~~~~~~~~~~~~~~~~~~~~\left.+
\frac{s+1}{s!(s-1)!}\Dd_{(\ad_s}\D_{(\a_{s-1}}\bar{\chi}_{\a(s-2)))\ad(s-1
)))}\right\}
\eea
%%%%
%%%%
\bea{l}
\frac{1}{s!(s-1)!}\D_{(\a_s}\Dd_{(\ad_{s-1}}G_{\a(s-1))\ad(s-2))}=\\
~~~~~=-\frac{2ic}{s!}\frac{s^2}{(2s+1)(s+1)}\pa_{(\a_s}{}^{\ad_s}
\left\{\vphantom{\frac{1}{2}}\Dd^2\D^{\g}H_{\g\a(s-1))\ad(s)}\right.
\IEEEyesnumber\\
~~~~~~~~~~~~~~~~~~~~~~~~~~~~~~~~~~~~~~~~~~~~+\frac{i(
s+1)}{s!}\Dd_{(\ad_s}\pa^{\g\gd}H_{\g\a(s-1))\gd\ad(s-1))}\\
~~~~~~~~~~~~~~~~~~~~~~~~~~~~~~~~~~~~~~~~~~~~\left.+
\frac{s+1}{s!(s-1)!}\Dd_{(\ad_s}\D_{(\a_{s-1}}\bar{\chi}_{\a(s-2)))
\ad(s-1))}\right\}\\
~~~~~~~-2ic\frac{s^2}{2s+1}\pa^{\a_{s+1}\ad_{s}}\left\{\frac{1}{(
s+1)!}\Dd^2\D_{(\a_{s+1}}H_{\a(s))\ad(s)}\right\}\\
~~~~~~~+\frac{2ic}{s!(s-1)!}\frac{s(s-1)}{2s+1}\pa_{(\a_s (\ad_{
s-1}}\left\{\vphantom{\frac{1}{2}}i\Dd^{\bd}\pa^{\g\gd}H_{\g\a(s-1))
\bd\gd\ad(s-2))}\right.\\
~~~~~~~~~~~~~~~~~~~~~~~~~~~~~~~~~~~~~~~~~~~~~~~~
\left.+\frac{1}{(s-1)!}\Dd^{\gd}\D_{(\a_{s-1}}\bar{\chi}_{\a(s-2)))
\gd\ad(s-2))}\right\}
\eea
%%%%
%%%%
\bea{l}
\frac{1}{(s-1)!}\Dd^{\ad_{s-1}}\D_{(\a_{s-1}}\bar{G}_{\a(s-2))\ad(
s-1)}=\\
~~~~~=-\frac{s}{s+1}\D^2G_{\a(s-1)\ad(s-2)}\\
~~~~~~+\frac{i}{(s-1)!}\frac{s(s-1)}{(s+1)^2}\pa_{(\a_{s-1}}{}^{\ad_{
s-1}}\bar{G}_{\a(s-2))\ad(s-1)}\\
~~~~~~-2ic\frac{s^2}{(2s+1)(s+1)}\pa^{\a_s\ad_{s-1}}\left\{\vphantom{
\frac{1}{2}}\D^2\Dd^{\gd}H_{\a(s)\gd\ad(s-1)}\right.\IEEEyesnumber\\
~~~~~~~~~~~~~~~~~~~~~~~~~~~~~~~~~~~~~~~~~~~~+\frac{i(s+1
)}{s!}\D_{(\a_s}\pa^{\g\gd}H_{\g\a(s-1))\gd\ad(s-1)}\\
~~~~~~~~~~~~~~~~~~~~~~~~~~~~~~~~~~~~~~~~~~~~\left.+\frac{
s+1}{s!(s-1)!}\D_{(\a_s}\Dd_{(\ad_{s-1}}\chi_{\a(s-1))\ad(s-2))}\right\}\\
~~~~~~~+2ic\frac{s(s-1)}{(s+1)^2}\frac{1}{(s-1)!}\pa_{(\a_{s-1}}{}^{
\ad_{s-1}}\left\{\vphantom{\frac{1}{2}}i\D^{\b}\pa^{\g\gd}H_{\b\g\a(
s-2))\gd\ad(s-1)}\right.\\
~~~~~~~~~~~~~~~~~~~~~~~~~~~~~~~~~~~~~~~~~~~~~~~~~~
\left.+\frac{1}{(s-1)!}\D^{\g}\Dd_{(\ad_{s-1}}\chi_{\g\a(s-2))\ad(s-2))}
\right\}
\eea

Let us define the following fields
\bea{l}
\frac{1}{(s+1)!}\Dd^2\D_{(\a_{s+1}}H_{\a(s))\ad(s))}|\equiv N_1~\psi_{
\a(s+1)\ad(s)}\\
\left\{\vphantom{\frac{1}{2}}\D^2\Dd^{\ad_s}H_{\a(s)\ad(s)}+\frac{i(s+1
)}{s!}\D_{(\a_s}\pa^{\g\gd}H_{\g\a(s-1))\gd\ad(s-1)}\right.\\
~~~~~~~~~~~~~~~~~~~~\left.+\frac{s+1}{s!(s-1)!}\D_{(\a_s}\Dd_{(\ad_{
s-1}}\chi_{\a(s-1))\ad(s-2))}\right\}|\equiv N_2~\psi_{\a(s)\ad(s-1)}\\
\left\{\vphantom{\frac{1}{2}}i\Dd^{\bd}\pa^{\g\gd}H_{\g\a(s-1)\bd\gd
\ad(s-2)}\right.\\
~~~\left.+\frac{1}{(s-1)!}\Dd^{\ad_{s-1}}\D_{(\a_{s-1}}\bar{\chi}_{\a(
s-2))\ad(s-1)}\right\}|\equiv N_3~\psi_{\a(s-1)\ad(s-2)}
\eea

Putting everything together, the component Lagrangian takes the form
\bea{ll}
\mathcal{L}_F=
&2ic|N_1|^2~\bar{\psi}^{\a(s)\ad(s+1)}\pa^{\a_{s+1}}{}_{\ad_{s+1}}
\psi_{\a(s+1)\ad(s)}\\
&-2ic\frac{s^2}{(2s+1)(s+1)}N_1N_2~\psi^{\a(s+1)\ad(s)}\pa_{\a_{
s+1}\ad_s}\psi_{\a(s)\ad(s-1)}+c.c.\\
&-2ic\frac{s^2}{(2s+1)(s+1)^2}|N_2|^2~\bar{\psi}^{\a(s-1)\ad(s)}\pa^{
\a_{s}}{}_{\ad_{s}}\psi_{\a(s)\ad(s-1)}\\
&-2ic\frac{s(s-1)}{(2s+1)(s+1)}N_2N_3~\psi^{\a(s)\ad(s-1)}\pa_{\a_{
s}\ad_{s-1}}\psi_{\a(s-1)\ad(s-2)}+c.c.\IEEEyesnumber\\
&-2ic\left(\frac{s-1}{s+1}\right)^2|N_3|^2~\bar{\psi}^{\a(s-2)\ad(s-1)}
\pa^{\a_{s-1}}{}_{\ad_{s-1}}\psi_{\a(s-1)\ad(s-2)}\\
&+\frac{1}{2c}\frac{(2s+1)(s-1)}{s^2(s+1)^2}G^{\a(s)\ad(s-1)}|\left(
\vphantom{\frac12}\D^2G_{\a(s-1)\ad(s-2)}\right.\\
&~~~~~~~~~~~~~~~~~~~~~~~~\left.-\frac{i}{2}\frac{s-1}{s+1}\frac{1
}{(s-1)!}\pa_{(\a_{s-1}}{}^{\ad_{s-1}}\bar{G}_{\a(s-2))\ad(s-1)}\right)|
+c.c\\
\eea

The last term in the Lagrangian is the algebraic kinetic energy term 
of two auxiliary fields and the rest of the terms are exactly the struc-
ture of a theory that describes helicity $h=s+1/2$. To have an exact 
match we choose coefficients
\bea{ll}
c=1~,~& N_2=-\frac{1}{\sqrt{2}}\frac{2s+1}{s}\\
N_1=\frac{1}{\sqrt{2}}~,~& N_3=\frac{1}{\sqrt{2}}\frac{s+1}{s-1}
\eea
So the fields that appear in the fermionic action are defined as:
\bea{l}
\rho_{\a(s-1)\ad(s-2)}\equiv G_{\a(s-1)\ad(s-2)}|\\
\b_{\a(s-1)\ad(s-2)}\equiv \left\{\vphantom{\frac12}\D^2G_{\a(s-1)\ad(
s-2)}\right.\\
~~~~~~~~~~~~~~~~~~~~\left.-\frac{i}{2}\frac{s-1}{s+1}\frac{1}{(s-1)!}
\pa_{(\a_{s-1}}{}^{\ad_{s-1}}\bar{G}_{\a(s-2))\ad(s-1)}\right\}|\\
\psi_{a(s+1)\ad(s)}\equiv\frac{\sqrt{2}}{(s+1)!}\Dd^2\D_{(\a_{s+1}}H_{
\a(s))\ad(s)}|\IEEEyesnumber\\
\psi_{\a(s)\ad(s-1)}\equiv-\sqrt{2}~\frac{s}{2s+1}\left\{\vphantom{\frac12}
\D^2\Dd^{\ad_s}H_{\a(s)\ad(s)}\right.\\
~~~~~~~~~~~~~~~~~~~~~~~~~~~~~~~~~+\frac{i(s+1)}{s!}\D_{(\a_s}
\pa^{\g\gd}H_{\g\a(s-1))\gd\ad(s-1)}\\
~~~~~~~~~~~~~~~~~~~~~~~~~~~~~~~~~\left.+\frac{s+1}{s!(s-1)!}
\D_{(\a_s}\Dd_{(\ad_{s-1}}\chi_{\a(s-1))\ad(s-2))}\right\}|\\
\psi_{\a(s-1)\ad(s-2)}\equiv\sqrt{2}~\frac{s-1}{s+1}\left\{\vphantom{\frac12}
i\Dd^{\bd}\pa^{\g\gd}H_{\g\a(s-1)\bd\gd\ad(s-2)}\right.\\
~~~~~~~~~~~~~~~~~~~~~~~~~~~~~~~~~~\left.+\frac{1}{(s-1)!}\Dd^{\ad_{
s-1}}\D_{(\a_{s-1}}\bar{\chi}_{\a(s-2))\ad(s-1)}\right\}|
\eea

The Lagrangian is
\bea{ll}
\mathcal{L}_F=&\rho^{\a(s)\ad(s-1)}\beta_{\a(s)\ad(s-1)}+c.c.\\
&+i~\bar{\psi}^{\a(s)\ad(s+1)}\pa^{\a_{s+1}}{}_{\ad_{s+1}}\psi_{\a(s+1)
\ad(s)}\\
&+i\left[\frac{s}{s+1}\right]~\psi^{\a(s+1)\ad(s)}\pa_{\a_{s+1}\ad_s}\psi_{
\a(s)\ad(s-1)}+c.c.\\
&-i\left[\frac{2s+1}{(s+1)^2}\right]~\bar{\psi}^{\a(s-1)\ad(s)}\pa^{\a_{s}}{
}_{\ad_{s}}\psi_{\a(s)\ad(s-1)}\IEEEyesnumber\\
&+i~\psi^{\a(s)\ad(s-1)}\pa_{\a_{s}\ad_{s-1}}\psi_{\a(s-1)\ad(s-2)}+c.c.\\
&-i~\bar{\psi}^{\a(s-2)\ad(s-1)}\pa^{\a_{s-1}}{}_{\ad_{s-1}}\psi_{\a(s-1)
\ad(s-2)}\\
\eea
and the gauge transformations of the fields are
\bea{ll}
\d_G\rho_{\a(s)\ad(s-1)}=0~,~&\d_G\psi_{\a(s+1)\ad(s)}=\frac{1}{s!(s+1)!}
\pa_{(\a_{s+1}(\ad_s}\xi_{\a(s))\ad(s-1))}\\
\d_G\b_{\a(s)\ad(s-1)}=0~,~&\d_G\psi_{\a(s)\ad(s-1)}=-\frac{1}{s!}\pa_{(
\a_{s}}{}^{\ad_s}\bar{\xi}_{\a(s-1))\ad(s)}\IEEEyesnumber\\
&\d_G\psi_{\a(s-1)\ad(s-2)}=\frac{s-1}{s}\pa^{\a_s\ad_{s-1}}\xi_{\a(s)\ad(
s-1)}\\
& \text{with } \xi_{\a(s)\ad(s-1)}=-i\sqrt{2}~\Dd^2L_{\a(s)\ad(s-1)}|
\eea
\subsubsection{Bosons}
~~ For the bosonic Lagrangian we do the same. The fields that appear 
in the action are defined as:
\bea{l}
A_{\a(s)\ad(s)}\equiv T_{\a(s))\ad(s)}|\\
U_{\a(s)\ad(s-2)}\equiv \frac{1}{s!}\D_{(\a_s}G_{\a(s-1))\ad(s-2)}|\\
u_{\a(s-1)\ad(s-1)}\equiv\frac{1}{2(s-1)!}\left\{\Dd_{(\ad_{s-1}}G_{\a(s-1
)\ad(s-2))}-\D_{(\a_{s-1}}\bar{G}_{\a(s-2))\ad(s-1)}\right\}|\\
v_{\a(s-1)\ad(s-1)}\equiv -\frac{i}{2(s-1)!}\left\{\Dd_{(\ad_{s-1}}G_{\a(s-1
)\ad(s-2))}+\D_{(\a_{s-1}}\bar{G}_{\a(s-2))\ad(s-1)}\right\}|\\
S_{\a(s-2)\ad(s-2)}\equiv\frac{1}{2}\left\{\D^{\a_{s-1}}G_{\a(s-1)\ad(s-2)}
+\Dd^{\ad_{s-1}}\bar{G}_{\a(s-2)\ad(s-1)}\right\}|\IEEEyesnumber\\
P_{\a(s-2)\ad(s-2)}\equiv -\frac{i}{2}\left\{\D^{\a_{s-1}}G_{\a(s-1)\ad(s-2)}
-\Dd^{\ad_{s-1}}\bar{G}_{\a(s-2)\ad(s-1)}\right\}|\\
h_{\a(s+1)\ad(s+1)}\equiv\frac{1}{2}\frac{1}{(s+1)!^2}\left[\D_{(\a_{s+1}},
\Dd_{(\ad_{s+1}}\right]H_{\a(s))\ad(s))}|\\
h_{\a(s-1)\ad(s-1)}\equiv -\frac{1}{2}\frac{s}{(2s+1)(s+1)^2}\left[\D^{\a_{
s}},\Dd^{\ad_{s}}\right]H_{\a(s)\ad(s)}|\\
~~~~~~~~~~~~~~~~~~~~-\frac{s}{(2s+1)(s+1)}\frac{1}{(s-1)!}\left(\vphantom
{\frac12} \D_{(\a_{s-1}}\bar{\chi}_{\a(s-2))\ad(s-1)}\right.\\
~~~~~~~~~~~~~~~~~~~~~~~~~~~~~~~~~~~~~~~~~~~~~~~~~~~~~~~
\left.-\vphantom{\frac12} \Dd_{(\ad_{s-1}}\chi_{\a(s-1)\ad(s-2))}\right)|
\eea
the gauge transformations are
\bea{l}
\d_G U_{\a(s)\ad(s-2)}=0,~~~~\d_G A_{\a(s)\ad(s)}=0\\
\d_G u_{\a(s-1)\ad(s-1)}=0,~~\d_G S_{\a(s-2)\ad(s-2)}=0
\IEEEyesnumber\\
\d_G v_{\a(s-1)\ad(s-1)}=0,~~\d_G P_{\a(s-2)\ad(s-2)}=0\\
\d_G h_{\a(s+1)\ad(s+1)}=\frac{1}{(s+1)!^2}\pa_{(\a_{s+1}(\ad_{s+1}}
\zeta_{\a(s))\ad(s))}\\
\d_G h_{\a(s-1)\ad(s-1)}=\frac{s}{(s+1)^2}\pa^{\a_{s}\ad_{s}}\zeta_{
\a(s)\ad(s)}\\
\eea
where
\bea{l}
\zeta_{\a(s)\ad(s)}=\frac{i}{2s!}\left(\D_{(\a_s}\bar{L}_{\a(s-1))\ad(s)}+
\Dd_{(\ad_s}L_{\a(s)\ad(s-1))}\right)|
\eea
and the bosonic Lagrangian is
\bea{ll}
\mathcal{L}_B=&-\frac{1}{4}\left[\frac{(2s+1)(s-1)}{s^2(s+1)}\right]~U^{
\a(s)\ad(s-2)}U_{\a(s)\ad(s-2)}+c.c.\\
&+\frac{1}{8}\left[\frac{2s+1}{s+1}\right]~A^{\a(s)\ad(s)}A_{\a(s)\ad(s)}\\
&-\frac{1}{2}\left[\frac{2s+1}{s^2}\right]~u^{\a(s-1)\ad(s-1)}u_{\a(s-1)\ad(
s-1)}\\
&-\frac{1}{2}\left[\frac{2s+1}{s^2}\right]~v^{\a(s-1)\ad(s-1)}v_{\a(s-1)\ad(
s-1)}\\
&-\frac{1}{2}\left[\frac{(2s+1)(s-1)^2}{s^3}\right]~S^{\a(s-2)\ad(s-2)}S_{\a(
s-2)\ad(s-2)}\\
&+\frac{1}{2}\left[\frac{(s-1)^2}{s^3}\right]~P^{\a(s-2)\ad(s-2)}P_{\a(s-2)
\ad(s-2)}\\
&+~h^{\a(s+1)\ad(s+1)}\Box h_{\a(s+1)\ad(s+1)}\\
&-\left[\frac{s+1}{2}\right]~h^{\a(s+1)\ad(s+1)}\pa_{\a_{s+1}\ad_{s+1}}
\pa^{\g\gd}h_{\g\a(s)\gd\ad(s)}\\
&+\left[s(s+1)\right]~h^{\a(s+1)\ad(s+1)}\pa_{\a_{s+1}\ad_{s+1}}\pa_{
\a_{s}\ad_{s}}h_{\a(s-1)\ad(s-1)}\\
&-\left[(s+1)(2s+1)\right]~h^{\a(s-1)\ad(s-1)}\Box h_{\a(s-1)\ad(s-1)}\\
&-\left[\frac{(s+1)(s-1)^2}{2}\right]~h^{\a(s-1)\ad(s-1)}\pa_{\a_{s-1}\ad_{
s-1}}\pa^{\g\gd}h_{\g\a(s-2)\gd\ad(s-2)}
\eea
and gives rise to the theory of  helicity $h=s+1$ as expected
\subsubsection{Off-shell degrees of freedom}
Let us count the bosonic degrees of freedom
\begin{center}
\begin{tabular}{|c|c|c|c|}
\hline 
\emph{fields} & \emph{d.o.f} & \emph{redundancy} & \emph{net} \\ 
\hline 
$h_{\a(s+1)\ad(s+1)}$ & $(s+2)^2$ & \multirow{2}{*}{$(s+1)^2$} & 
\multirow{2}{*}{$s^2+2s+3$} \\ 
\cline{1-2}
$h_{\a(s-1)\ad(s-1)}$ & $s^2$ & & \\ 
\hline 
$u_{\a(s-1)\ad(s-1)}$ & $s^2$ & 0 & $s^2$  \\ 
\hline 
$v_{\a(s-1)\ad(s-1)}$ & $s^2$ & 0 & $s^2$ \\ 
\hline 
$A_{\a(s)\ad(s)}$ & $(s+1)^2$ & 0 & $(s+1)^2$ \\ 
\hline 
$U_{\a(s)\ad(s-2)}$ & $2(s+1)(s-1)$ & 0 & $2(s+1)(s-1)$ \\ 
\hline 
$S_{\a(s-2)\ad(s-2)}$ & $(s-1)^2$ & 0 & $(s-1)^2$ \\ 
\hline 
$P_{\a(s-2)\ad(s-2)}$ & $(s-1)^2$ & 0 & $(s-1)^2$ \\ 
\hline\hline 
\multicolumn{2}{c|}{} & \emph{Total} & $8s^2+4$ \\ 
\cline{3-4}
\end{tabular}
\end{center}
and the same counting for the fermionic degrees of freedom
\begin{center}
\begin{tabular}{|c|c|c|c|}
\hline 
\emph{fields} & \emph{d.o.f} & \emph{redundancy} & \emph{net} \\ 
\hline 
$\psi_{\a(s+1)\ad(s)}$ & $2(s+2)(s+1)$ & \multirow{3}{*}{$2(s+1)s$} 
& \multirow{3}{*}{$4s^2+4s+4$} \\ 
\cline{1-2}
$\psi_{\a(s)\ad(s-1)}$ & $2(s+1)s$ & & \\ 
\cline{1-2}
$\psi_{\a(s-1)\ad(s-2)}$ & $2s(s-1)$ & & \\ 
\hline 
$\rho_{\a(s-1)\ad(s-2)}$ & $2(s-1)s$ & 0 & $2(s-1)s$ \\ 
\hline 
$\b_{\a(s-1)\ad(s-2)}$ & $2(s-1)s$ & 0 & $2(s-)s$ \\ 
\hline\hline 
\multicolumn{2}{c|}{} & \emph{Total} & $8s^2+4$ \\ 
\cline{3-4}
\end{tabular}
\end{center}

\subsubsection{SUSY-transformation laws}
~~ The explicit expressions for the SUSY-transformation laws 
of the fields can be found in the same way as for case (I).  For 
the fermionic fields:
\bea{ll}
\d_S\rho_{\a(s-1)\ad(s-2)}=&-\e^{\a_s}U_{\a(s)\ad(s-2)}
\IEEEyesnumber\\
&+\left[\frac{s-1}{s}\right]\frac{1}{(s-1)!}\e_{(\a_{s-1}}\left[S_{
\a(s-2))\ad(s-2)}+iP_{\a(s-2))\ad(s-2)}\right]\\
&-\ed^{\ad_{s-1}}\left[u_{\a(s-1)\ad(s-1)}+iv_{\a(s-1)\ad(s-1)}
\right]
\eea
\bea{l}
\d_S\b_{\a(s-1)\ad(s-2)}=\\
=\frac{i}{2}\frac{s^2}{s+1}\ed^{\ad_{s-1}}\pa^{\a_s\ad_s}
A_{\a(s)\ad(s)}\\
~+\frac{s^2}{2s+1}\ed^{\ad_{s-1}}\pa^{\a_{s+1}\ad_{s+1}}
\pa^{\a_s\ad_s}h_{\a(s+1)\ad(s+1)}\\
~-2s\ed^{\ad_{s-1}}\Box h_{\a(s-1)\ad(s-2)}\\
~-\frac{s(s-1)^2}{2s+1}\frac{1}{(s-1)!^2}\ed^{\ad_{s-1}}\pa_{(
\a_{s-1}(\ad_{s-1}}\pa^{\b\bd}h_{\b\a(s-2))\bd\ad(s-2))}\\
~-\frac{i}{(s-1)!}\ed^{\ad_{s-1}}\pa^{\a_s}{}_{(\ad_{s-1}}U_{\a(
s)\ad(s-2))}\\
~+\frac{s-2}{s-1}\frac{i}{(s-2)!}\ed_{(\ad_{s-2}}\pa^{\b\bd}U_{\b\a(
s-1)\bd\ad(s-3))}\\
~+\frac{1}{2}\frac{s-1}{s+1}\frac{i}{(s-1)!}\ed^{\ad_{s-1}}\pa_{(\a_{
s-1}}{}^{\ad_s}\bar{U}_{\a(s-2))\ad(s)}\IEEEyesnumber\\
~+\frac{(s-1)(2s^2+2s+1)}{2s(s+1)}\frac{i}{(s-1)!^2}\ed^{\ad_{s-1}}
\pa_{(\a_{s-1}(\ad_{s-1}}S_{\a(s-2))\ad(s-2))}\\
~-\frac{(s-1)(2s^2+4s+3)}{2s(s+1)(2s+1)}\frac{1}{(s-1)!^2}\ed^{\ad_{
s-1}}\pa_{(\a_{s-1}(\ad_{s-1}}P_{\a(s-2))\ad(s-2))}\\
~-\frac{(s-2)(3s+2)}{2s(s+1)}\frac{i}{(s-2)!(s-1)!}\ed_{(\ad_{s-2}}\pa_{
(\a_{s-1}}{}^{\bd}S_{\a(s-2))\bd\ad(s-3))}\\
~+\frac{(s-2)(s+2)}{2s(s+1)}\frac{1}{(s-2)!(s-1)!}\ed_{(\ad_{s-2}}\pa_{
(\a_{s-1}}{}^{\bd}P_{\a(s-2))\bd\ad(s-3))}\\
~-\frac{1}{2}\frac{s-1}{s+1}\frac{i}{(s-1)!}\e^{\b}\pa_{(\a_{s-1}}{}^{\ad_{
s-1}}u_{\b\a(s-2))\ad(s-1)}\\
~-\frac{1}{2}\frac{s-1}{s+1}\frac{1}{(s-1)!}\e^{\b}\pa_{(\a_{s-1}}{}^{\ad_{
s-1}}v_{\b\a(s-2))\ad(s-1)}
\eea
\bea{ll}
\d_S\psi_{\a(s+1)\ad(s)}=&\frac{\sqrt{2}i}{(s+2)!}\e^{\a_{s+2}}\pa_{(\a_{
s+2}}{}^{\ad_{s+1}}h_{\a(s+1))\ad(s+1)}\\
&-\frac{1}{\sqrt{2}}\frac{s}{s+2}\frac{i}{(s+1)!}\e_{(\a_{s+1}}\pa^{\g\gd}
h_{\g\a(s))\gd\ad(s)}\IEEEyesnumber\\
&+\frac{1}{2\sqrt{2}}\frac{2s+1}{s+1}\frac{1}{(s+1)!}\e_{(\a_{s+1}}
A_{\a(s))\ad(s)}
\eea
\bea{ll}
\d_S\psi_{\a(s)\ad(s-1)}=&\frac{1}{\sqrt{2}}\frac{s+1}{s}\frac{1}{s!}\e_{(
\a_s}\left[-u_{\a(s-1))\ad(s-1)+iv_{\a(s-1))\ad(s-1)}}\right]\\
&+\frac{1}{\sqrt{2}}\frac{s-1}{s}\frac{1}{(s-1)!}\ed_{(\ad_{s-1}}U_{\a(s)\ad(
s-2))}\\
&-\frac{1}{2\sqrt{2}}\frac{s}{s+1}\ed^{\ad_s}A_{\a(s)\ad(s)}
\IEEEyesnumber\\
&-\frac{is}{\sqrt{2}}\ed^{\ad_s}\pa^{\a_{s+1}\ad_{s+1}}h_{\a(s+1)\ad(
s+1)}\\
&+\frac{is(s+2)}{\sqrt{2}s!s!}\ed^{\ad_{s}}\pa_{(\a_s(\ad_s}h_{\a(s-1))
\ad(s-1))}
\eea
\bea{ll}
\d_S\psi_{\a(s-1)\ad(s-2)}=&-\frac{1}{\sqrt{2}}\frac{(2s+1)(s-1)}{s^2(s+1
)}\ed^{\ad_{s-1}}u_{\a(s-1)\ad(s-1)}\\
&-\frac{i}{\sqrt{2}}\frac{(2s+1)(s-1)}{s^2(s+1)}\ed^{\ad_{s-1}}v_{\a(s-1)
\ad(s-1)}\\
&-\frac{1}{\sqrt{2}}\frac{(s-1)^2(2s+1)}{s^2(s+1)}\frac{1}{(s-1)!}\e_{\a_{
s-1}}S_{\a(s-2))\ad(s-2)}\IEEEyesnumber\\
&+\frac{i}{\sqrt{2}}\frac{(s-1)^2}{s^2(s+1)}\frac{1}{(s-1)!}\e_{\a_{s-1}}
P_{\a(s-2))\ad(s-2)}\\
&+i\sqrt{2}\frac{(s-1)^2(s+1)}{s}\frac{1}{(s-1)!}\e_{(\a_{s-1}}\pa^{\g\gd}
h_{\g\a(s-2))\gd\ad(s-2)}
\eea
and the SUSY-transformation laws for the bosonic fields are:
\bea{ll}
\d_S A_{\a(s)\ad(s)}=&-\frac{i\sqrt{2}}{(s+1)!}\ed^{\ad_{s+1}}\pa^{\a_{
s+1}}{}_{(\ad_{s+1}}\psi_{\a(s+1)\ad(s))}+c.c.\\
&+\frac{i\sqrt{2}}{s!}\frac{s^2}{(s+1)(2s+1)}\ed_{(\ad_s}\pa^{\g\gd}\psi_{
\g\a(s)\gd\ad(s-1))}+c.c.\\
&+\frac{i\sqrt{2}}{(s+1)!s!}\frac{s}{s+1}\ed^{\ad_{s+1}}\pa_{(\a_s(\ad_{
s+1}}\bar{\psi}_{\a(s-1))\ad(s))}+c.c.\\
&+\frac{i\sqrt{2}}{s!s!}\frac{s}{(s+1)^2}\ed_{(\ad_s}\pa_{(\a_s}{}^{\gd}
\bar{\psi}_{\a(s-1))\gd\ad(s-1))}+c.c.\IEEEyesnumber\\
&-\frac{i\sqrt{2}}{s!s!}\frac{s}{2s+1}\ed_{(\ad_s}\pa_{(\a_s\ad_{s-1}}
\psi_{\a(s-1))\ad(s-2))}+c.c.\\
&-\frac{i}{s!s!}\frac{s-1}{s+1}\ed_{(\ad_s}\pa_{(\a_s\ad_{s-1}}\rho_{
\a(s-1))\ad(s-2))}+c.c.
\eea
\bea{ll}
\d_S U_{\a(s)\ad(s-2)}=&\frac{1}{s!}\e_{(\a_s}\b_{\a(s-1))\ad(s-2)}\\
&-\frac{i}{s!(s-1)!}\ed^{\ad_{s-1}}\pa_{(\a_s(\ad_{s-1}}\rho_{\a(s-1))
\ad(s-2))}\\
&+\frac{i}{s!(s-2)!}\frac{s-2}{s-1}\ed_{(\ad_{s-2}}\pa_{(\a_s}{}^{\gd}
\rho_{\a(s-1))\gd\ad(s-3))}\\
&+\frac{i}{s!}\frac{s-1}{2(s+1)}\e_{(\a_s}\pa_{\a_{s-1}}{}^{\ad_{s-1}}
\bar{\rho}_{\a(s-2))\ad(s-1)}\\
&-i\sqrt{2}\frac{s^2}{2s+1}\ed^{\ad_{s-1}}\pa^{\a_{s+1}\ad_s}\psi_{
\a(s+1)\ad(s)}\IEEEyesnumber\\
&-\frac{i\sqrt{2}}{s!}\frac{s}{s+1}\ed^{\ad_{s-1}}\pa_{(\a_s}{}^{\ad_s}
\bar{\psi}_{\a(s-1))\ad(s)}\\
&+\frac{i\sqrt{2}}{s!(s-1)!}\frac{s(s+1)}{2s+1}\ed^{\ad_{s-1}}\pa_{(
\a_s(\ad_{s-1}}\psi_{\a(s-1))\ad(s-2))}
\eea

\bea{l}
\d_S\left(u_{\a(s-1)\ad(s-1)}+iv_{\a(s-1)\ad(s-1)}\right)=\\
~~~~~=i\sqrt{2}\frac{s^2}{2s+1}\e^{\a_s}\pa^{\a_{s+1}\ad_s}\psi_{
\a(s+1)\ad(s)}\\
~~~~~~+i\sqrt{2}\frac{s}{s+1}\frac{1}{s!}\e^{\a_s}\pa_{(\a_s}{}^{\ad_s
}\bar{\psi}_{\a(s-1))\ad(s)}\\
~~~~~~-i\sqrt{2}\frac{s(s+1)}{2s+1}\frac{1}{s!(s-1)!}\e^{\a_s}\pa_{(
\a_s(\ad_{s-1}}\psi_{\a(s-1))\ad(s-2)))}\IEEEyesnumber\\
~~~~~~-\frac{s^2}{(s+1)(s-1)}\frac{1}{(s-1)!}\e_{(\a_{s-1}}\bar{\b}_{\a(
s-2))\ad(s-1)}\\
~~~~~~-i\sqrt{2}\frac{s^2}{(s+1)(s-1)}\frac{1}{(s-1)!}\e_{(\a_{s-1}}\pa^{
\g\ad_s}\bar{\psi}_{\g\a(s-2))\ad(s)}\\
~~~~~~+i\sqrt{2}\frac{s^2}{(s+1)(s-1)}\frac{1}{(s-1)!}\e_{(\a_{s-1}}\pa^{
\g}{}_{(\ad_{s-1}}\psi_{\g\a(s-2))\ad(s-2))}\\
\eea
\bea{l}
\d_S\left(S_{\a(s-2)\ad(s-2)}+iP_{\a(s-2)\ad(s-2)}\right)=\\
~~~~=\e^{\a_{s-1}}\b_{\a(s-1)\ad(s-2)}\\
~~~~~+\frac{i}{2}\frac{s-1}{s+1}\frac{1}{(s-1)!}\e^{\a_{s-1}}\pa_{(\a_{
s-1}}{}^{\ad_{s-1}}\bar{\rho}_{\a(s-2))\ad_{s-1}}\\
~~~~~-\frac{i}{(s-1)!}\ed^{\ad_{s-1}}\pa^{\a_{s-1}}{}_{(\ad_{s-1}}\rho_{
\a(s-1)\ad(s-2))}\\
~~~~~+\frac{i}{(s-2)!}\frac{s-2}{s-1}\ed_{(\ad_{s-2}}\pa^{\g\gd}\rho_{\g\a
(s-2)\gd\ad(s-3))}\IEEEyesnumber\\
~~~~~-\frac{s}{s+1}\ed^{\ad_{s-1}}\bar{\b}_{\a(s-2)\ad(s-1)}\\
~~~~~-i\sqrt{2}\frac{s}{s+1}\ed^{\ad_{s-1}}\pa^{\a_{s-1}\ad_{s}}\bar{\psi
}_{\a(s-1)\ad(s)}\\
~~~~~+\frac{i\sqrt{2}}{(s-1)!}\frac{s}{s+1}\ed^{\ad_{s-1}}\pa^{\a_{s-1}}{}_{
(\ad_{s-1}}\psi_{\a(s-1)\ad(s-2))}
\eea
\bea{ll}
\d_S h_{\a(s+1)\ad(s+1)}=&\frac{1}{\sqrt{2}(s+1)!}\e_{(\a_{s+1}}\bar{
\psi}_{\a(s))\ad(s+1)}+c.c.\IEEEyesnumber
\eea
\bea{ll}
\d_S h_{\a(s-1)\ad(s-1)}=&\frac{1}{\sqrt{2}}\frac{1}{(s+1)^2}\e^{\a_s}
\psi_{\a(s)\ad(s-1)}+c.c.\\
&+\frac{1}{\sqrt{2}}\frac{1}{(s+1)}\frac{1}{(s-1)!}\e_{(\a_{s-1}}\bar{\psi
}_{\a(s-2))\ad(s-1)}+c.c.\IEEEyesnumber\\
&-\frac{1}{2}\frac{s-1}{s(s+1)^2}\frac{1}{(s-1)!}\e_{(\a_{s-1}}\bar{\rho}_{
\a(s-2))\ad(s-1)}+c.c.
\eea

\section{Hints for $\mathcal{N}=2$}

The massless irreducible representations of $4D$, $\mathcal{N}=2$ 
Super-Poincar\'{e} group for super-helicity $Y$ describe helicities $
\lambda=Y+1,~\lambda=Y+1/2,~\lambda=Y+1/2,~\lambda=Y$.  At 
least on-shell that looks like the direct sum of two $\mathcal{N}=1$ 
massless irreducible representations, one describing super-helicity 
$Y+1/2$ and the other one describing super-helicity $Y$. Therefore 
one will be tempted to try to combine the theory of integer super-helicity 
$Y=s$ presented in \cite{GKS} with one of the theories of half-integer 
super-helicity $Y=s+1/2$ presented here in order to construct an $
\mathcal{N}=2$ representation. The question is which pair [integer,
half-integer(I)] or [integer,half-integer(II)] will be the one to give the 
$\mathcal{N}=2$ representation. In an attempt to find the answer the 
authors of \cite{GKS}, by trial and error concluded that the answer was 
[integer,half-integer(I)].

The counting of the degrees of freedom argument provides a very simple 
explanation why this is the case. The integer theory has exactly the same 
degrees of freedom as the half-integer (I) theory. This is a sign that if we 
add together the two theories then in principle we can have a second 
direction of supersymmetry that will map the bosons (fermions) of one
theory to the fermions (bosons) of the other theory. This can only happen 
if the number of bosons and fermions match exactly, as they do. Therefore 
we can construct an irreducible representation of $4D,~\mathcal{N}=2$ 
Super-Poincar\'{e} group. Also in the same manner we can understand 
why a possible pair of integer theory with half-integer (II) theory can never 
work.

\section{Summary}
~~ We continue the programm started in \cite{IntSpin}
for the case of half-integer super-helicity. There are two classes of theories 
that describe the same physical system but they will turn out to have different 
off-shell structure. We reproduce the superspace action for both of them in 
terms of unconstrained superfields, following the redundancy guideline as 
was force on us by the representation theory of the Super-Poincar\'{e} group.

For Transverse theories this action is a representative of a bigger two 
parameter family of actions that are all equivalent and they are related 
by superfield redefinitions. That is not the case for Longitudinal theories, 
were the action is unique and no local redefinitions of the superfields can 
be done.

Finally, using the equations of motion generated by the superspace action 
we define the components of the theory. We derive the component action 
in \emph{diagonal} form and calculate the susy transformation laws for each 
one of them. A counting of the off-shell degrees of freedom for transverse 
theories will give the same number as the theory of integer super-helicity 
and therefore explains why they can be combined to give an $\mathcal{N
}=2$ irreducible representation. The same counting for Longitudinal theories 
will also prove that 1) they can not be used together with integer super-helicity 
theories to make $\mathcal{N}=2$ representations and 2) Off-shell Longitudinal 
theories are not equivalent to Transverse theories since there can be no 1-1
mapping between the two off-shell..

\section*{Acknowledgements}
~~ K. Koutrolikos wants to thank Dr. W.D. Linch and Dr. K. Stiffler 
for useful comments and discussions. This research has been 
supported in part by NSF Grant  PHY-09-68854, the J.~S. Toll 
Professorship endowment and the UMCP Center for String \& 
Particle Theory.

\end{document}